\documentclass[prb,twocolumn,showpacs,superscriptaddress]{revtex4}
\usepackage[dvips]{epsfig}
\usepackage{graphicx}
\usepackage{amsmath,amssymb,lscape,float}
\usepackage{bm}
\usepackage{bbm}
\usepackage{dsfont}

\newcommand{\norm}[1]{\lVert#1\rVert}
\renewcommand{\Re}{\mathop{\rm Re}\nolimits}
\renewcommand{\Im}{\mathop{\rm Im}\nolimits}
\newcommand{\hc}{\mathop{\rm H.c.}\nolimits}

\newcommand{\dg}{\dagger}

\newcommand{\gpsi}{{\psi}}

\newcommand{\gphi}{{\phi}}

\newcommand{\gvphi}{{\varphi}}

\newcommand{\mbf}{\mathbf}
\renewcommand{\vec}{\mbf}
\newcommand{\veck}{\vec k}
\newcommand{\vecr}{\vec r}

\makeatletter
\renewcommand\appendix{\par
  \setcounter{section}{0}%
  \setcounter{subsection}{0}%
  \setcounter{equation}{0}%
  \gdef\thesection{\Alph{section}}
  \@addtoreset {equation}{section}
  \renewcommand{\theequation}{\thesection\arabic{equation}}}
\makeatother

\begin{document}
\title{Polaronic signatures and spectral properties\\
       of graphene antidot lattices}

\author{Vladimir M. Stojanovi\'c}
\email{vladimir.stojanovic@unibas.ch}
\affiliation{Department of Physics, University of Basel,
 Klingelbergstrasse 82, CH-4056 Basel, Switzerland}
 
\author{Nenad Vukmirovi\'c}
\affiliation{Computational Research Division,
Lawrence Berkeley National Laboratory,
Berkeley, California 94720, USA} 

\author{C. Bruder}
\affiliation{Department of Physics, University of Basel,
 Klingelbergstrasse 82, CH-4056 Basel, Switzerland}

\date{\today}
\begin{abstract}
We explore the consequences of electron-phonon (e-ph) coupling in  
graphene antidot lattices (graphene nanomeshes), i.e., triangular superlattices 
of circular holes (antidots) in a graphene sheet. They display a direct band gap whose 
magnitude can be controlled via the antidot size and density. The relevant coupling mechanism 
in these semiconducting counterparts of graphene is the modulation of the nearest-neighbor 
electronic hopping integrals due to lattice distortions (Peierls-type e-ph coupling).  
We compute the full momentum dependence of the e-ph vertex functions for a number 
of representative antidot lattices. Based on the latter, we discuss the origins of
the previously found large conduction-band quasiparticle spectral weight due to e-ph coupling. 
In addition, we study the nonzero-momentum quasiparticle properties with the aid of the self-consistent 
Born approximation, yielding results that can be compared with future angle-resolved photoemission 
spectroscopy measurements. Our principal finding is a significant e-ph mass enhancement, an 
indication of polaronic behavior. This can be ascribed to the peculiar momentum dependence of the
e-ph interaction in these narrow-band systems, which favors small phonon momentum scattering. 
We also discuss implications of our study for recently fabricated large-period
graphene antidot lattices.
\end{abstract}
\maketitle
\section{Introduction}
The discovery of freestanding graphene has spawned huge interest in the electronic and transport 
properties of this material.~\cite{GrapheneReviews} In particular, a great deal of 
research effort is presently being dedicated to graphene-based superlattices.~\cite{MoireSuperlattice,BerkeleySuperlatt,PedersenAntidot:08,GrapheneSuperlattPRB,Balog+:10} 
Among them, a class of superhoneycomb systems~\cite{Shima+Aoki:93} -- {\em 
graphene antidot lattices} ({\em graphene nanomeshes}~\cite{Bai+:10}) -- was 
introduced~\cite{PedersenAntidot:08} by analogy with the concept of antidot lattices 
defined atop a two-dimensional electron gas in a semiconductor heterostructure.~\cite{SemiconAntidot} 
The electronic,~\cite{AntidotExper,Vanevic+:09,Fuerst++:09} optical,~\cite{PedersenOptical} 
and magnetic~\cite{KineziAntidot} properties of these superlattices result from a subtle 
interplay of the intrinsic features of graphene and a lattice periodicity imposed by holes
(antidots) in a graphene sheet.

Aside from some fundamental aspects,~\cite{Tworzydlo:06} the main incentive behind the
current graphene-related research activity stems from the prospect of a carbon-based
electronics.~\cite{CarbonElectronics} Graphene displays exceptional properties, 
such as room-temperature ballistic transport~\cite{Mueller+:09} on a 
submicron scale with intrinsic mobility as high as $200,000$ cm$^{2}$/Vs and the 
possibility of heavy doping without altering significantly the charge-carrier mobility.
Yet, intrinsic semimetallic graphene is of limited utility for fabricating electronic devices~\cite{CarbonElectronics} 
because the transmission probability of Dirac electrons across a potential barrier 
is always unity -- independent of the height and width of the barrier -- a manifestation 
of Klein tunneling.~\cite{KleinTunneling} As a result, the conductivity cannot be controlled 
through a gate voltage, which, however, is a prerequisite for operation of a conventional field-effect transistor 
where switching is achieved by gate-voltage induced depletion. Thus it is crucial to open a band gap in graphene.
A possible approach entails processing graphene sheets into nanoribbons, which show a confinement-induced gap that scales
inversely with their width. To get a gap sufficient for room-temperature 
transistor operation ($E_{\rm g}\sim 0.5$\:eV) one needs ribbons with a width of only few nanometers.  

Graphene antidot lattices have a direct electronic band gap stemming
from the quantum confinement brought about by the periodic 
potential of a regular arrangement of antidots. The magnitude of the gap 
follows a generic scaling relation~\cite{KineziAntidot} -- a linear increase 
with the product of the antidot size and density -- and can thus be externally controlled. 
It is this tunability of the band gap -- not available when a gap is induced by 
growing graphene epitaxially on substrates such as SiC -- that makes  
these superlattices interesting from both the fundamental and practical viewpoints. 
They can be thought of as highly interconnected networks of nanoribbons, 
and can in principle be fabricated on suspended graphene monolayers via 
electron-beam lithography, whose current resolution limit is as low as $10$\:nm.
Yet, being serial in nature, this method is not applicable to large-area
patterning of graphene. An alternative fabrication approach, block copolymer 
lithography, has recently been utilized to fabricate very uniform, free-standing triangular graphene 
antidot lattices.~\cite{Bai+:10,Kim+:10} Superior room-temperature electrical properties of the 
fabricated devices -- compared to graphene nanoribbons -- have also been demonstrated.~\cite{Bai+:10}

A thorough understanding of e-ph scattering is of utmost importance because 
the latter determines the ultimate performance limit of any electronic 
device: in the high-bias regime of transport the e-ph scattering increases the differential
resistance. In the present work, we address the e-ph coupling~\cite{EPHgeneralMarsiglio} 
in graphene antidot lattices and explore some of the ensuing physical consequences. 
The major e-ph-interaction mechanism in this system, 
as well as in $\pi$-electron systems in general,~\cite{SSH,MahanEtAl,NonlocalCouplingLong} 
is the modulation of electronic hopping integrals due to lattice 
distortions (Peierls-type e-ph coupling~\cite{PeierlsCoupling}). 

It is by now widely accepted that e-ph coupling in graphene is comparatively
weak:~\cite{ParkZgraphene:07,GraphenePhononEffect} for instance, the angle-resolved 
photoemission spectroscopy (ARPES) data on inelastic carrier lifetime~\cite{Bostwick++:07} 
were consistently explained without even invoking phonon-related 
effects.~\cite{Hwang++:07} Here we demonstrate that in graphene antidot lattices, 
which have completely different band structure than their ``parent material'', 
e-ph coupling is very significant, i.e., properties of these systems 
undergo a strong phonon-induced renormalization. We ascribe this not only to their 
narrow-band character and low dimensionality, but also to the peculiar momentum-dependence 
of the e-ph interaction, favoring small phonon momentum scattering. 

Unlike in our recent work,~\cite{Vukmirovic+:10} which was solely concerned 
with the zero-momentum conduction-band quasiparticle spectral weight for different 
graphene antidot lattices, in the present work we put an additional emphasis on the 
directional dependence of the e-ph mass enhancement. This is particularly appropriate 
because of the anisotropic (in momentum space) character of the relevant e-ph coupling.
We also describe the nonzero-momentum spectral properties due to the e-ph 
coupling at the level of the self-consistent Born approximation. In this way, we obtain
results that can be compared with future ARPES measurements.
 
The outline of the paper is as follows. In Sec.~\ref{antidotlatt} we introduce the system 
under consideration, along with the notation and conventions to be used throughout. 
In subsequent Secs.~\ref{bandstruc} and \ref{phonspect}
we briefly recapitulate the main features of the band structures and phonon spectra
of graphene antidot lattices, respectively. Sec.~\ref{ephcoupled} is set aside for
a detailed account of the microscopic mechanism of e-ph coupling in the system, with emphasis 
on its resulting momentum dependence. The following Sec.~\ref{PhononRenormalize} is devoted 
to the effect that this momentum-dependence has on the phonon-induced renormalization 
of electronic properties. In Sec.~\ref{discuss} we discuss some implications of our 
study for optical absorption experiments and charge transport, as well as
possible future work. We conclude, with a brief summary of the paper, in 
Sec.~\ref{sumconc}. Some mathematical details are deferred to 
Appendices \ref{vertfuncapp} and \ref{renormdirect}.

\section{Notation and conventions}  \label{antidotlatt}
\begin{figure}[t!]
\begin{center}
\epsfig{file=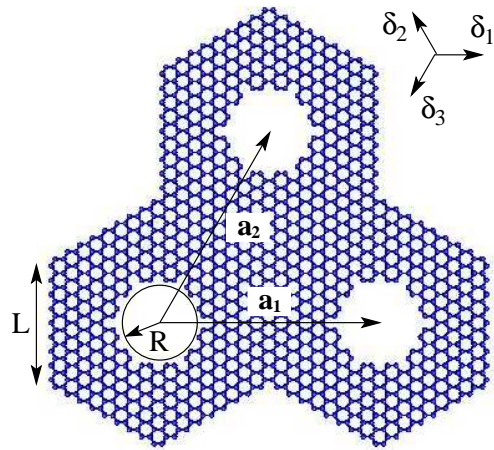,width=0.75\linewidth,clip=} 
\caption{(Color online) A finite segment of the graphene antidot lattice $\{L,R\}$ 
with circular antidots. Here $\mathbf{a}_{1}$ and $\mathbf{a}_{2}$ are the superlattice 
basis vectors; $L$ and $R$ are, respectively, the side length of the hexagonal unit cell
and the antidot radius, both expressed in units of the graphene lattice constant 
$a\approx 2.46\:$\AA. 
The vectors $\bm{\delta}_{1},\:\bm{\delta}_{2}$, and $\bm{\delta}_{3}$ 
specify positions of the nearest neighbors of a carbon atom on sublattice $A$.} \label{AntidotFig}
\end{center}
\end{figure}

Triangular graphene antidot lattices with circular antidots
are illustrated in Fig.~\ref{AntidotFig}. Their geometry is uniquely specified 
by two dimensionless parameters: the side length of their hexagonal unit cells  
($L$) and the antidot radius ($R$), both expressed in units of the graphene lattice 
constant $a\approx 2.46\:$\AA. (Note that $a=a_{cc}\sqrt{3}$, where $a_{cc}=1.42\:$\AA\: 
is the distance between nearest-neighbor carbon atoms.) 
While $L$ is invariably an integer, and defines the superlattice period 
$La\sqrt{3}$ (center-to-center distance between nearest-neighbor antidots), 
$R$ can also take non-integer values. Therefore, the notation $\{L,R\}$
will hereafter be used to specify different antidot lattices. 
If a carbon atom (henceforth abbreviated as C atom) on sublattice A is taken to be 
the origin, the adjacent C atoms are determined by the vectors 
$\bm{\delta}_{1}$, $\bm{\delta}_{2}$, and $\bm{\delta}_{3}$ (see Fig.~\ref{AntidotFig}). 
(For an origin at sublattice B, the corresponding vectors would be 
$-\bm{\delta}_{1}$,$-\bm{\delta}_{2}$,$-\bm{\delta}_{3}$.)

In what follows, the vectors $\mathbf{R}$ will designate the unit cells 
($N$ of them) of an antidot lattice, while $\mathbf{d}_{m}$ ($m=1,\ldots,N_{\textrm{at}}$) 
will specify the positions of the C atoms within a unit cell. 
Thus the vectors $\mathbf{R}+\mathbf{d}_{m}$ uniquely specify the positions of all C atoms.

\section{Band structure of graphene antidot lattices} \label{bandstruc}
Much like some other types of graphene-based superlattices,~\cite{MoireSuperlattice}
typical antidot lattices contain extremely large numbers of C atoms
in their unit cells; in particular, here we consider superlattices with 
$N_{\textrm{at}}\sim 300-1600$ atoms per unit cell. Such a large system size
prohibits the use of methods based on the density functional theory (DFT) for 
the band-structure and phonon calculations. Instead, we model the band structure using a 
nearest-neighbor $\pi$-orbital tight-binding model Hamiltonian
\begin{equation}\label{tbHam}
\hat{H}_{\textrm{e}}=-\frac{t}{2}\sum_{\mathbf{R},m,\bm{\delta}}
\big(\hat{a}^\dg_{\mathbf{R}+\mathbf{d}_{m}+\bm{\delta}}
\hat{a}_{\mathbf{R}+\mathbf{d}_{m}}+\hc\big) \:,
\end{equation}
\noindent where $\hat{a}_{\mathbf{R}+\mathbf{d}_{m}}$ 
($\hat{a}^{\dagger}_{\mathbf{R}+\mathbf{d}_{m}}$)
destroys (creates) an electron in the $2p_z$ orbital
$\gvphi(\vec r-\mbf R -\mbf d_m)$ of the C atom located at
the position $\mathbf{R}+\mathbf{d}_{m}$, $\bm{\delta}$ stands for
the nearest neighbors of that C atom, and $t\approx 2.8$\:eV is 
the nearest-neighbor hopping integral. The tight-binding method is 
known to be capable of reproducing very accurately the low-energy part of the DFT band structure 
of graphene.~\cite{GrapheneReviews} Its accuracy in the case of graphene antidot lattices,
i.e., its good agreement with the DFT results for lattices with very small unit 
cells, has recently been demonstrated.~\cite{Fuerst++:09}
 
The Bloch wave functions corresponding to the energy eigenvalues
$\varepsilon_{n}(\veck)$, where $n$ is the band index and $\mbf k$ is 
a quasimomentum in the Brillouin zone (BZ), are given by
\begin{equation} \label{BlochWF}
\gpsi_{n\mbf k}(\mbf r)=\sum_m C_{m}^{n,\veck}\gphi_{m\veck}(\vecr) \:,
\end{equation}
\noindent where $\gphi_{m\veck}(\vecr)$ is the discrete Fourier transform of 
$\varphi(\mathbf{r}-\mathbf{R}-\mathbf{d}_{\scriptscriptstyle m})$.
Due to the sixfold-rotational point-group symmetry of the system, 
the coefficients $C^{n,\mbf k}_{m}$ obey the conditions
\begin{equation}\label{Csymm}
C_{m}^{n,\veck}=e^{i\theta(\veck)}C_{m'}^{n,\veck'} \:,
\end{equation}
\noindent where $\theta(\veck)$ is an irrelevant $\veck$-dependent phase and
$\veck'$ ($m'$) is obtained by rotating the vector $\mathbf{k}$ (atom $m$) through
an angle of $\pi/3$.

Unlike graphene itself, graphene antidot lattices show a band structure characteristic of 
direct band gap semiconductors. In particular, the lattices with circular antidots, that 
we are concerned with here, fall into the class of superhoneycomb systems (with a 
unit cell that must have at least the six-fold rotational symmetry) whose possible 
band structures were classified by Shima and Aoki.~\cite{Shima+Aoki:93} 

In what follows, we study the $\{L,R=5\}$ ($9\leq L\leq 19$) and $\{L,R=7\}$ 
($12\leq L\leq 20$) families of graphene antidot lattices. We find that these  
lattices are extreme narrow-band systems: for instance, in the $\{L,R=5\}$ family 
the conduction-electron band width $W_{c}$ is in the range $0.11-0.14$\:eV; 
in the $\{L,R=7\}$ family $W_{c}$ is even smaller ($0.02-0.04$\:eV). The band gaps,
unlike band widths, decrease with $L$ for given $R$. In the $\{L,R=5\}$ family, they are 
in the range between $E_{\rm g}=0.146$\:eV (for $L=19$) and $E_{\rm g}=0.735$\:eV (for $L=9$);
in the $\{L,R=7\}$ family, the corresponding range is between $E_{\rm g}=0.118$\:eV (for $L=20$)
and $E_{\rm g}=0.301$\:eV (for $L=12$). The band gap scaling with antidot dimensions
was studied in Refs.~\onlinecite{PedersenOptical} and \onlinecite{KineziAntidot}.  

It is worth noting that as a consequence of the bipartite nature of the 
underlying honeycomb lattice, the resulting energy spectrum of our nearest-neighbor 
tight-binding model displays particle-hole symmetry.~\cite{Vanevic+:09} This property 
is not perfectly retained in the exact band structure.~\cite{Fuerst++:09}

\section{Phonon spectra of graphene antidot lattices} \label{phonspect}
The phonon spectrum of graphene was studied extensively, using either 
DFT methods or effective models.~\cite{Saito+:98,GraphenePhononCalcL,Zimmermann+:08,Perebeinos+Tersoff:09}
The effective models are either based on approximating directly the
interatomic force constants, with terms that describe coupling of atoms up to some 
maximum distance, or on adopting an analytic expression for the interaction 
energy of two or more C atoms. In either case, they
contain adjustable parameters whose values are determined by fitting 
available experimental- and/or DFT data. A well-known class of 
empirical interatomic potentials for carbon are the Tersoff-Brenner 
potentials.~\cite{Brenner+Tersoff} However, these potentials, with 
interatomic-interaction range that corresponds only to second-nearest-neighbor 
distance, do not provide the accuracy typically needed in phonon 
calculations.~\cite{GraphenePhononCalcL,Zimmermann+:08} 
A generalization of these potentials, with interactions up to fourth-nearest-neighbor, 
was implemented by Tewary and Yang for calculating the phonon spectra and elastic 
constants of graphene.~\cite{GraphenePhononCalcL} 
Yet, this graphene-specific generalization does not lend itself to the use in other 
carbon-based systems, as the generalized potentials become unstable away from the 
equilibrium lattice configuration of graphene.

In the present work, we determine the phonon spectra of the $\{L,R=5\}$ 
and $\{L,R=7\}$ graphene antidot lattices (with the values of $L$ specified above)
using two independent models that have recently been shown to yield very accurate 
results for graphene itself: the fourth-nearest-neighbor force-constant (4NNFC) 
method~\cite{Jishi4nnfc:93} in the parametrization of 
Zimmermann {\em et al.},~\cite{Zimmermann+:08} and the valence force-field (VFF) model, 
developed for graphene by Perebeinos and Tersoff.~\cite{Perebeinos+Tersoff:09} These two models
belong to the two aforementioned mutually distinct groups of effective lattice-dynamical models.
Both models are, in principle, applicable to any $sp^{2}$-hybridized carbon system.

The 4NNFC model, introduced by Saito {\em et al.}~\cite{Saito+:98}, 
entails a direct parametrization of the real-space force constants up to fourth-order 
neighbors, hence the name. The model includes a set of $12$ adjustable parameters, 
whose values are determined by fitting the {\it ab-initio} phonon dispersion 
(for the parameter values, see Table I in Ref.~\onlinecite{Zimmermann+:08}). These 
force-constant parameters correspond to the radial (bond-stretching), and both 
in-plane and out-of-plane tangential (bond-bending) directions between an atom and its
$n$-th nearest neighbors. The underlying formalism is described in detail
in Ref.~\onlinecite{CNTbook}.

The VFF model of Perebeinos and Tersoff is based on an explicit expression for the 
interaction energy, which includes six different contributions, respectively related 
to: bond-stretching, bending, out-of-plane vibrations, misalignment 
of neighboring $\pi$ orbitals, bond-order, and coupling between bond stretching and 
bond bending [see Eqs.~(1) and (2) in Ref.~\onlinecite{Perebeinos+Tersoff:09}]. Each 
contribution is parameterized by the corresponding stiffness constant;
their values (see Table~I in Ref.~\onlinecite{Perebeinos+Tersoff:09}) 
are determined by fitting experimental and {\em ab-initio} data for phonon 
energies and elastic constants. The model makes no reference to any underlying 
crystal structure; the only restriction to its use is that the local geometry be consistent 
with $sp^{2}$ bonding, i.e., that the three adjacent C atoms are not too far from 
being $120^{\circ}$ apart.~\cite{Perebeinos+Tersoff:09} 
Thus the model can be directly applied not only to graphene 
but also to various other allotropes of carbon, with graphene antidot lattices being 
an example of such structures. 

In each particular case, we first find the equilibrium lattice configuration 
by relaxing the atoms until the forces on all of them are smaller than 
$10^{-5}$ eV/\AA. We then construct the force-constant tensor
\begin{equation}
D_{m\beta, m'\beta'} 
(\mbf{R}-\mbf{R}')\equiv\frac{\partial^2 E_{\textrm{tot}}(\{\mbf{u}_{m}(\mbf{R})\})}
{\partial u_{m\beta}(\mbf{R})\partial u_{m'\beta'}(\mbf{R}')} \:,
\end{equation}
\noindent where $u_{m\beta}(\mathbf{R})$ is the displacement -- in direction 
$\beta$ ($\beta=x,y,z$) from the equilibrium position -- of an atom at 
$\mathbf{R}+\mbf d_m$, and $E_{\textrm{tot}}(\{\mbf{u}_{m}(\mbf{R})\})$ 
the total lattice potential energy. The normal-mode frequencies 
$\omega_{\lambda}(\mathbf{q})$ and eigenvectors 
$\mathbf{v}^{\lambda}(\mathbf{q})=[\mathbf{v}_{1}^{\lambda}
(\mathbf{q}),\ldots,\mathbf{v}_{N_{\textrm{at}}}^{\lambda}(\mathbf{q})]^{T}$ 
(with $\lambda=1,\ldots,3N_{\textrm{at}}$ enumerating phonon branches) are 
obtained by solving numerically the secular equation 
for the dynamical matrix $\mbf{D}(\mathbf{q})\equiv\sum_{\mbf{R}}
\mbf{D}(\mbf{R})\:e^{-i\mathbf{q}\cdot\mathbf{R}}$.

The salient feature of the computed phonon spectra is that the highest
optical-phonon energy at $\mathbf{q}=0$ is essentially inherited from graphene 
itself and thus only weakly dependent on $L$ and $R$. Generally speaking, the 
comparatively high optical phonon energies in carbon-based 
systems (graphene, nanotubes, intramolecular modes in fullerides) -- which 
typically extend up to about $200$\:meV -- are related to the 
low atomic mass of carbon and to the stiffness of the C-C bond.
The energies we obtain for this highest-lying mode in antidot lattices are 
around $195.3$\:meV in the 4NNFC approach and $197.5$\:meV in the VFF approach. 
The highest-energy phonon modes have in common that they do not involve significant 
displacements in the vicinity of the antidot edges.
\begin{figure}[t!]
\hspace*{-7.5mm}
\epsfig{file=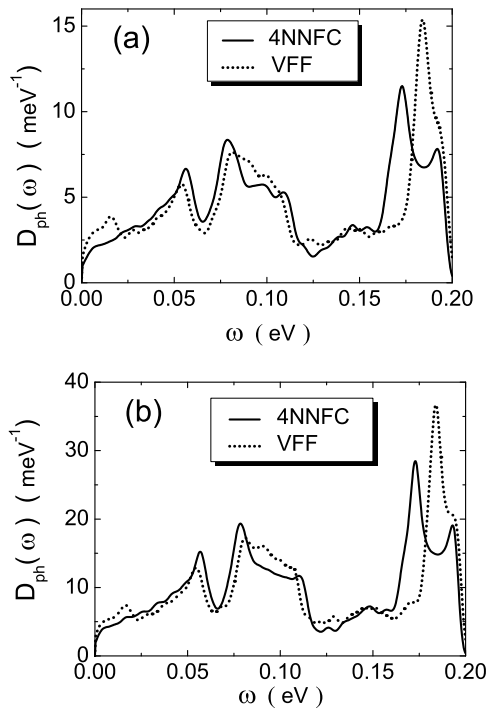,width=0.75\linewidth,clip=}
\caption{\label{PhononDOS} The phonon density-of-states for
the (a) $\{L=9,R=5\}$ and (b) $\{L=13,R=7\}$ graphene antidot lattices, 
obtained using the 4NNFC (solid line) and VFF (dashed line) methods.}
\end{figure}

Given the number of phonon modes involved, 
the most meaningful way of comparing the results of the two methods 
employed entails calculation of the phonon density-of-states 
\begin{equation}
D_{\textrm{ph}}(\omega)\equiv \frac{1}{N}\sum_{\mathbf{q},\lambda}
\delta[\omega-\omega_{\lambda}(\mathbf{q})] \:.
\end{equation}
\noindent This quantity is normalized here such that when integrated
over the whole range of $\omega$ it yields the total number of the degrees 
of freedom per unit cell, i.e., the number of phonon branches $3N_{\textrm{at}}$.
In particular, Figs.~\ref{PhononDOS}(a) and \ref{PhononDOS}(b) illustrate 
such a comparison for the $\{L=9,R=5\}$ and $\{L=13,R=7\}$ antidot lattices, respectively. 
The plots show the qualitative agreement between the two approaches, as well as
very similar character of the phonon density-of-states in both families of
antidot lattices studied. The accuracy of both methods can potentially be improved 
as system-specific experimental data become available, allowing to determine more 
accurately the values of adjustable parameters.

\section{Electron-phonon coupling in graphene antidot lattices} \label{ephcoupled}
The dominant e-ph coupling mechanism in $\pi$-electron systems is the modulation 
of hopping integrals due to lattice distortions,~\cite{MahanEtAl} i.e., Peierls-type 
coupling.~\cite{PeierlsCoupling} The latter forms the basis of the 
Su-Schrieffer-Heeger (SSH) model.~\cite{SSH,SSHnanotubes} In $sp^{2}$-bonded carbon-based 
systems optical phonons modify the length of the in-plane $\sigma$ bond between two 
C atoms, thus altering the overlap between the out-of-plane $\pi$ orbitals centered 
on these atoms. As a result, the $\pi$-electron hopping integrals are dynamically 
bondlength-dependent; to a linear approximation, they are proportional to the 
bondlength modulation, with $\alpha=5.27\:\textrm{eV/\AA}$ being
the corresponding e-ph coupling constant.~\cite{MahanEtAl,Vukmirovic+:10} 
The coupling to acoustic phonons, on the other hand, is known to be rather weak 
in graphene and carbon nanotubes.~\cite{ParkZgraphene:07,Borysenko++:10} 
In our system, where acoustic phonons have even lower energies than in 
graphene and carbon nanotubes, this coupling is expected 
to be even weaker and can therefore be safely neglected.

The momentum-space form of the e-ph-coupling Hamiltonian reads
\begin{equation}\label{mscoupling}
\hat{H}_{\textrm{ep}}=\frac{1}{\sqrt{N}}
\sum_{\mathbf{k,q},\lambda,nn'}
\gamma_{nn'}^{\lambda}(\mathbf{k,q}) \:
\hat{a}_{n,\mathbf{k+q}}^{\dagger}
\hat{a}_{n',\mathbf{k}}(\hat{b}_{-\mathbf{q},\lambda}
^{\dagger}+\hat{b}_{\mathbf{q},\lambda}) \:,
\end{equation}
\noindent where $\hat{a}_{n,\mathbf{k}}$ ($\hat{a}_{n,\mathbf{k}}^{\dagger}$) destroys 
(creates) an electron with quasimomentum $\mbf k$ in the $n$-th Bloch band, and
$\gamma_{nn'}^{\lambda}(\mathbf{k,q})$ stands for the (bare) e-ph interaction 
vertex function. Quite generally, $\gamma_{nn'}^{\lambda}(\mathbf{k,q})$ has the meaning
of the matrix element -- between the electronic Bloch states 
$(n',\mathbf{k})$ and $(n,\mathbf{k+q})$ -- of the induced self-consistent potential 
$\Delta U_{\lambda\mbf q}$ per unit displacement along the phonon normal coordinate 
that corresponds to the mode $\lambda\mbf q$:
\begin{equation} 
\gamma_{nn'}^{\lambda}(\mathbf{k,q})\equiv\frac{1}
{\sqrt{2M\omega_{\lambda}(\mathbf{q})}}
\:\langle n',\mathbf{k}\:|\Delta 
U_{\lambda\mbf q}|\:n,\mathbf{k+q}\rangle\:.
\end{equation}
\noindent In systems where $\Delta U_{\lambda\mbf q}$ can be determined as a by-product of
a DFT band-structure calculation (in which case it represents a derivative of
the self-consistent Kohn-Sham potential), like for graphene,~\cite{ParkZgraphene:07,Borysenko++:10} 
the last expression incorporates contributions of the phonon mode $\lambda\mathbf{q}$ to all 
possible e-ph coupling mechanisms. Hence there is no need to assume a 
particular microscopic e-ph interaction form. Here, however, where such an approach 
is unfeasible, evaluation of $\gamma_{nn'}^{\lambda}(\mathbf{k,q})$ is based on the adopted real-space 
form of the e-ph coupling written in the tight-binding electron basis. 
In the case at hand
\begin{equation} 
\gamma_{nn'}^{\lambda}(\mathbf{k,q})=
V_{nn'}^\lambda (\mbf k ,\mbf q)+W_{nn'}^\lambda(\mbf k ,\mbf q)\:,
\end{equation}
\noindent where $V_{nn'}^\lambda (\mbf k ,\mbf q)$ and 
$W_{nn'}^\lambda(\mbf k ,\mbf q)$ are the respective contributions of 
the neighbors within a single unit cell and those from adjacent unit cells.
As shown explicitly in Appendix \ref{vertfuncapp}, 
\begin{multline}\label{vertex1}
V_{nn'}^\lambda (\mbf k ,\mbf q)=\frac{\alpha}{\sqrt{8M\omega_\lambda(\mbf q)}}
\sum_{m,\bm{\delta}}\bar{\bm{\delta}}\cdot [\mbf v^\lambda_{m+\delta}
(\mbf q)-\mbf v^\lambda_m(\mbf q)]\\
\times\big[(C^{n, \mbf k+\mbf q}_{m+\delta})^* C^{n',\mbf k}_m+ 
(C^{n, \mbf k + \mbf q}_m)^* C^{n', \mbf k}_{m+\delta}\big] \:,
\end{multline}
\noindent where $\bar{\bm{\delta}}\equiv\bm{\delta}/\norm{\bm{\delta}}$, while $m+\delta$ 
denotes neighbors $\mathbf{d}_{m}+\bm{\delta}$ of site $\mathbf{d}_{m}$, and
\begin{multline}\label{vertex2}
W_{nn'}^\lambda (\mbf k ,\mbf q)=\frac{\alpha}{\sqrt{8M\omega_\lambda(\mbf q)}}
\sideset{}{'}\sum_{m,\bm{\delta},\mbf{a}}\bar{\bm{\delta}}
\cdot[e^{i\mbf q\cdot\mbf a}\mbf v^\lambda_{m_1}(\mbf q)-\mbf v^\lambda_m(\mbf q)]\\
\times\big[ e^{-i(\mbf k+\mbf q)\cdot \mbf a}  
(C^{n, \mbf k+\mbf q}_{m_1})^* C^{n',\mbf k}_m+ e^{i\mbf k\cdot \mbf a}  
(C^{n, \mbf k + \mbf q}_m)^* C^{n', \mbf k}_{m_1}\big]\:.
\end{multline}
\noindent The prime on the last sum indicates that the sum is restricted to
the neighbors $\mbf d_m+\bm{\delta}$ of site $\mbf d_m$ that satisfy 
the condition $\mbf d_m + \bm \delta =\mbf a + \mbf d_{m_1}$
for some $m_1=m_1(\bm{\delta})$, with $\mbf a = \pm \mbf a_1$, $\pm \mbf a_2$, 
$\pm(\mbf a_1-\mbf a_2)$ (see Fig.~\ref{AntidotFig}). 

For the conduction band ($n\rightarrow c$) electrons, in the absence of interband 
($n\neq n'$) scattering, the e-ph Hamiltonian of Eq.~\eqref{mscoupling}
reduces to the effective form
\begin{equation}\label{cbcoupling}
\hat{H}^{(\textrm{c})}_{\textrm{ep}}=\frac{1}{\sqrt{N}}\sum_{\mathbf{k,q},\lambda}
\gamma_{\textrm{cc}}^{\lambda}(\mathbf{k,q}) \:
\hat{a}_{\textrm{c},\mathbf{k+q}}^{\dagger}\hat{a}_{\textrm{c},\mathbf{k}}
(\hat{b}_{-\mathbf{q},\lambda}^{\dagger}+\hat{b}_{\mathbf{q},\lambda}) \:.
\end{equation}
Evaluation of the e-ph vertex functions $\gamma_{\textrm{cc}}^{\lambda}(\mathbf{k,q})$ 
based on Eqs.~\eqref{vertex1} and \eqref{vertex2} requires
the full knowledge of the phonon polarization vectors $\mbf v^\lambda_m(\mbf q)$ 
and Bloch wave functions, with the latter ones entering through the coefficients $C_{m}^{n=c,\veck}$.

\section{Phonon-induced renormalization} \label{PhononRenormalize}
\subsection{Consequences of the momentum dependence of e-ph coupling} \label{PhononRenormA}
At variance with most other familiar types of e-ph interaction,
which are either completely momentum independent (Holstein-type coupling~\cite{Holstein:59})
or only weakly phonon-momentum dependent (Fr\"{o}hlich-type coupling~\cite{DevreeseAlexandrovReview}), 
Peierls-type coupling depends strongly on both the electron and phonon 
momenta. This translates into a strongly-retarded nature of the interaction, which can have
a variety of physical consequences.~\cite{zolicouple} Due to particular geometry of the 
system, the momentum dependence of e-ph interaction in the case at hand is clearly more complicated 
than that of the standard SSH coupling, which is usually studied on monoatomic
lattices (one-dimensional, or a two-dimensional square one).

The quantity that characterizes phonon-induced renormalization for conduction-band
electrons is the quasiparticle spectral weight due to e-ph coupling, 
$Z_{c}(\mathbf{k})\equiv|\langle\Psi_{c\mathbf{k}}|\psi_{c\mathbf{k}}\rangle|^{2}$\:,
where $|\psi_{c\mathbf{k}}\rangle\equiv\hat{a}^{\dagger}_{c\mathbf{k}}|0\rangle_{\textrm{e}}$ 
is the bare conduction-electron Bloch state (i.e., a common eigenstate of $\hat{H}_{\textrm{e}}$ 
and the total electron momentum operator $\hat{\mathbf{K}}_{\textrm{e}}\equiv\sum_{n',\mathbf{k'}}
\mathbf{k'}\:\hat{a}^{\dagger}_{n',\mathbf{k'}}\hat{a}_{n'\mathbf{k'}}$) and 
$|\Psi_{c\mathbf{k}}\rangle$ that of the coupled e-ph system
(i.e., an eigenstate of the full Hamiltonian of the coupled e-ph system
and the total crystal momentum operator $\hat{\mathbf{K}}\equiv\sum_{n',\mathbf{k'}}
\mathbf{k'}\:\hat{a}^{\dagger}_{n',\mathbf{k'}}\hat{a}_{n',\mathbf{k'}}+\sum_{\lambda,\mathbf{q'}}
\mathbf{q'}\:\hat{b}^{\dagger}_{\lambda,\mathbf{q'}}\hat{b}_{\lambda,\mathbf{q'}}$).

In a recent work,~\cite{Vukmirovic+:10} we evaluated $Z_{\textrm{c}}(0)$ for 
the $\{L,R=5\}$ and $\{L,R=7\}$ families of lattices, finding a rather strong 
phonon-induced renormalization $Z_{\textrm{c}}(0)=0.20-0.27$ [i.e., $Z_{\textrm{c}}^{-1}(0)=3.7-5$],
compared to graphene~\cite{GraphenePhononEffect} where $Z=0.93$ at Dirac points, 
or larger.~\cite{ParkZgraphene:07} These results, with an excellent agreement between the 4NNFC 
and VFF approaches, seem to suggest that charge carriers acquire polaronic character. In what 
follows, we analyze in detail the most important phonon modes and
momentum dependence of their corresponding e-ph vertex functions.

The largest contribution to the spectral weight ($75-80$ percent in total) 
comes from the high-energy modes, more precisely two groups of modes centered 
around $173$ meV and $194$ meV (cf. Fig.~\ref{PhononDOS}). 
These modes have mixed character, being neither purely transverse nor 
longitudinal. We also note that within the phonon energy intervals that 
contribute to the spectral weight, the contribution is dominated by a few 
modes only. For example, in the case of $\{L=13,R=5\}$ [$\{L=15,R=7\}$] lattice 
only $9$ ($24$) modes are sufficient to provide already $50$ percent of the 
spectral weight.

\begin{figure}[t!]
%\hspace*{-7.5mm}
\epsfig{file=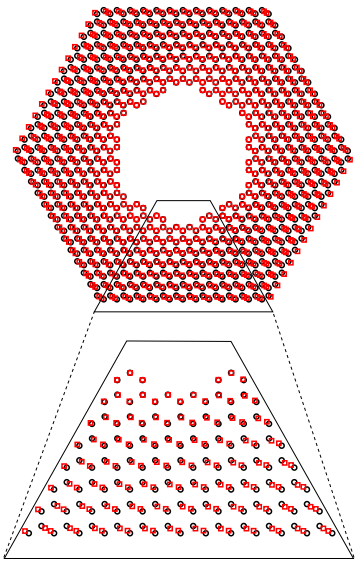,width=0.55\linewidth,clip=}
\caption{\label{AtomDisp1}(Color online) The atomic displacement pattern 
corresponding to the zone-center phonon mode at energy $E_{\lambda}\approx 197.5$\:meV 
of the $\{L=13,R=5\}$ graphene antidot lattice (top) and its zoomed version (bottom). 
The displaced (undisplaced) atomic positions are designated by the squares (dark circles).}
\end{figure}

\begin{figure}[t!]
%\hspace*{-7.5mm}
\epsfig{file=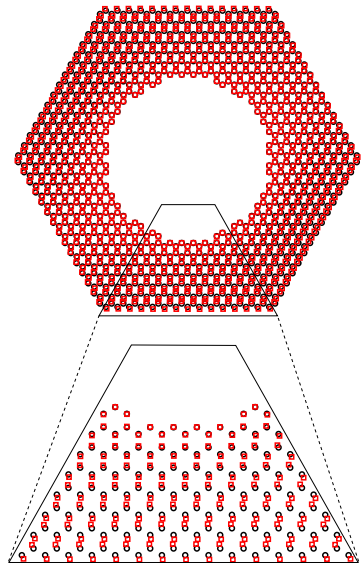,width=0.55\linewidth,clip=}
\caption{\label{AtomDisp2}(Color online) The atomic displacement pattern 
corresponding to the zone-center phonon mode at energy $E_{\lambda}\approx 197.5$\:meV 
of the $\{L=15,R=7\}$ graphene antidot lattice (top) and its zoomed version (bottom). 
The displaced (undisplaced) atomic positions are designated by the squares (dark circles).}
\end{figure}

The highest-energy phonon branches yield the largest individual contribution 
to $Z_{\textrm{c}}^{-1}(0)$ in all the lattices studied; these contributions are 
in the range $0.57-0.65$ in the $\{L,R=5\}$ lattices and $0.27-0.34$ in the $\{L,R=7\}$ 
lattices. Examples of the atomic displacement patterns of such modes for $\{L=13,R=5\}$ 
and $\{L=15,R=7\}$ antidot lattices are given in Figs.~\ref{AtomDisp1} and \ref{AtomDisp2}, from 
which we can infer that these modes do not involve significant atomic displacements 
in the vicinity of the antidot edges, but only away from them. This is a consequence
of the fact that the system is ``stiffer'' to the displacements of atoms away from edges (which have
three neighbors) than those along the edges (only two neighbors). Similar argument explains
the dominance of tangential displacements (over radial ones) in the highest-energy phonon
modes of fullerides.~\cite{Gunnarsson:97}

The last conclusion about the importance of the highest-energy optical 
phonon branch is similar to that in graphene: based on a first-principles
study of the e-ph interaction in graphene, Park {\em et al.}~\cite{ParkZgraphene:07}
concluded that the results could be reproduced by assuming a single Einstein
mode at around $200$\:meV. 

On the other hand, the optical phonons at energies below $30$ meV 
are far less important than their high-energy counterparts, with 
their cumulative contribution to the overall spectral weight being at 
most $20$ percent in the antidot lattices studied. This can  already be understood 
based on the shape of the phonon density-of-states (Fig.~\ref{PhononDOS}), 
from which it is obvious that in the low-energy region the phonon density-of-states 
is comparatively small.

\begin{figure}[b!]
%\hspace*{-7.5mm}
\epsfig{file=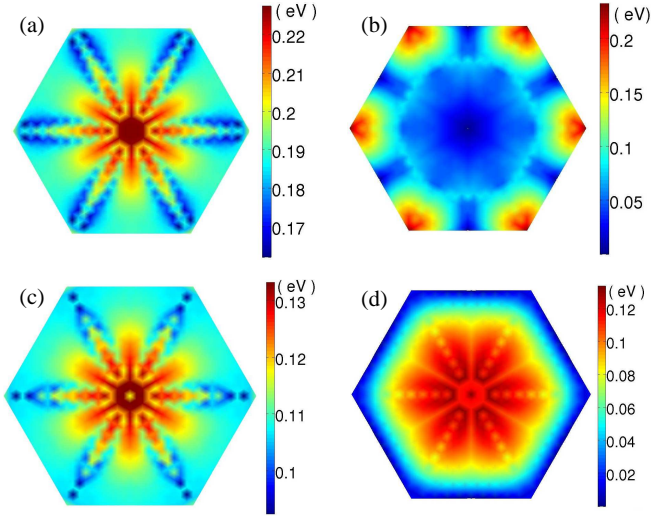,width=0.997\linewidth,clip=}
\caption{\label{VertexFunc}(Color online) The $\mbf q$-dependence of the moduli   
$|\gamma^{\lambda}_{\textrm{cc}}(\mbf{k}=0,\mbf{q})|$ of the bare e-ph vertex functions, 
for selected high-energy phonon branches $\lambda$, in the hexagonal Brillouin zones of the 
$\{L=13,R=5\}$ [(a)-(b)] and $\{L=15,R=7\}$ [(c)-(d)] graphene antidot lattices. The 
energies of the corresponding optical phonons at $\mbf{q}=0$, as obtained 
using the VFF approach, are $E_{\lambda}\approx 197.5$\:meV [(a),(c),(d)]
and $E_{\lambda}\approx 197.4$\:meV [(b)].}
\end{figure}

In order to get a qualitative insight into the obtained results, it is instructive
to analyze in detail the underlying e-ph coupling mechanism.
The momentum dependence of the bare vertex functions for an electron at the
conduction-band bottom ($\mbf{k}=0$) and the dominant phonon modes in typical 
lattices is depicted in Figs.~\ref{VertexFunc}(a)-(d). What can be inferred from 
the plots is that the e-ph coupling has a very rich momentum dependence, characterized 
by a high degree of anisotropy within the irreducible wedge of the BZ. We find several 
characteristic momentum-space profiles for the coupling functions, some of which correspond 
to e-ph coupling peaked at (or around) $\mbf q=0$ [Figs.~\ref{VertexFunc}(a),(c) and (d)], 
and the remaining ones to the cases where the latter have maxima at nonzero phonon 
momenta [Fig.~\ref{VertexFunc}(b)]. 

In all the cases considered, the coupling to the highest-energy phonon branch is 
strongest at zero phonon momentum and in its immediate vicinity. The last observation 
leads us to conclude that the small-phonon-momentum ($\mbf{q}\approx 0$) scattering 
involving this mode is to a large extent responsible for the sizeable phonon-induced renormalization
that we have obtained. As can be inferred by analyzing the expression for $Z_{\textrm{c}}^{-1}(0)$ 
in the lowest-order Rayleigh-Schr\"{o}dinger perturbation theory [Eq.~(8) in Ref.~\onlinecite{Vukmirovic+:10}],
the $\mbf{q}\approx 0$ terms yield the largest contribution to the relevant integral since for small 
$\mbf{q}$ values the denominator of the integrand is very small. The last assertion about 
the relevance of $\mbf{q}\approx 0$ scattering is quite general and holds 
even when the e-ph interaction is completely momentum independent 
(local in real space), as is the case for Holstein-type interaction. 
However, our system constitutes an extreme realization of such a physical scenario as the 
function $|\gamma^{\lambda}_{\textrm{cc}}(\mbf{k}=0,\mbf{q})|^{2}$, corresponding to the 
highest-lying phonon, attains its largest values precisely at $\mbf{q}=0$ and thus leads
to a large mass enhancement. Needless to say, distinct examples of momentum-dependent 
e-ph interactions do exist where the physical situation is rather different; 
for the SSH coupling~\cite{SSH,zolicouple} 
$\gamma_{\scriptscriptstyle\textrm{SSH}}(\mbf{k},\mbf{q})\propto \sin(\mbf{k}\cdot\mbf{a})
-\sin\big[(\mbf{k+q})\cdot\mbf{a}\big]$ one has 
$\gamma_{\scriptscriptstyle\textrm{SSH}}(\mbf{k}=0,\mbf{q})\propto|\mbf{q}|$ at small $\mbf{q}$. 
Thus the SSH coupling for $\mbf{k}=0$ vanishes as $\mbf{q}\rightarrow 0$. 
Another familiar example is furnished by the coupling of Zhang-Rice singlets 
to the ``breathing'' (oxygen bond-stretching) phonon modes in cuprates,~\cite{Gunnarsson+Roesch} 
where $\gamma(\mbf{k},\mbf{q})=\gamma(\mbf{q})\propto \sqrt{\sin^{2}(q_{x}/2)+\sin^{2}(q_{y}/2)}$ 
and is therefore also suppressed at small phonon momenta.

\subsection{Nonzero-momentum spectral properties due to e-ph coupling}
In the following, we study nonzero-momentum spectral properties
due to e-ph coupling. Because of a considerable computational burden involved,
we perform calculations on selected antidot lattices -- $\{L=9,R=5\}$ and $\{L=13,R=5\}$ 
from the $\{L,R=5\}$ family, as well as $\{L=15,R=7\}$ from the $\{L,R=7\}$
family -- within the VFF approach.

\begin{figure}[t!]
\hspace*{-7.5mm}
\epsfig{file=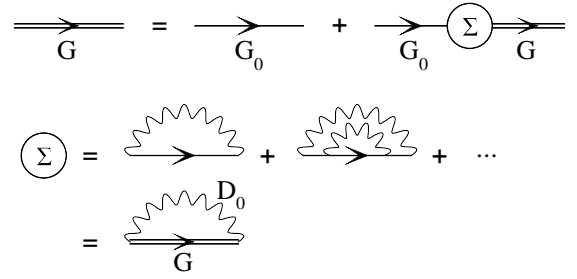,width=0.85\linewidth,clip=}
\caption{\label{FeynDiag} Pictorial representation of the Dyson equation (top).
Illustration of the SCBA self-energy, given by a sum over all noncrossing
diagrams (bottom).}
\end{figure}

We determine the conduction-band electron self-energy using the 
self-consistent Born approximation (SCBA).~\cite{Gunnarsson+Roesch} 
In the SCBA the electron self-energy is obtained by summing over 
all the non-crossing diagrams, as illustrated in Fig.~\ref{FeynDiag}. 
Although this approximation neglects e-ph vertex corrections 
in the self-energy calculation, it yields good results for not-too-strong 
e-ph coupling.~\cite{Vukmirovic+:07}
The SCBA self-energy is given by
\begin{multline}\label{selfen}
\Sigma_{\textrm{c}}(\mathbf{k},\omega)=\frac{i}{2\pi N}
\sum_{\mathbf{q},\lambda}\int d\Omega\:|\gamma^{\lambda}_{\textrm{cc}}
(\mathbf{k},\mathbf{q})|^{2}\\
\times D_{\lambda}(\mathbf{q},\Omega)\:G_{\textrm{c}}(\mathbf{k+q},\omega-\Omega) \:, 
\end{multline}
\noindent where
\begin{eqnarray}
G_{\textrm{c}}(\mathbf{k},\omega)&\equiv & \frac{1}{\omega-
\varepsilon_{\textrm{c}}(\mathbf{k})-\Sigma_{\textrm{c}}
(\mathbf{k},\omega)+i\delta} \label{SCBAelprop} \:, \\
D_{\lambda}(\mathbf{q},\Omega)&\equiv & \frac{1}
{\Omega-\omega_{\lambda} +i0^{+}} - \frac{1}
{\Omega+\omega_{\lambda}-i0^{+}}\:,
\end{eqnarray}
\noindent are the interacting electron and the free phonon propagators, 
respectively. The frequency integration in Eq.~\eqref{selfen} can be 
carried out explicitly, leading to
\begin{multline}\label{}
\Sigma_{\textrm{c}}(\mathbf{k},\omega)=\frac{1}{N}\sum_{\mathbf{q},\lambda}
\:|\gamma^{\lambda}_{\textrm{cc}}(\mathbf{k},\mathbf{q})|^{2} \\
\times\frac{1}{\omega-\omega_{\lambda}-\varepsilon_{\textrm{c}}(\mathbf{k}+\mathbf{q})-
\Sigma_{\textrm{c}}(\mathbf{k}+\mathbf{q},\omega-\omega_{\lambda})+i\delta} \:.
\end{multline}

\begin{figure}[b!]
\hspace*{-7.5mm}
\epsfig{file=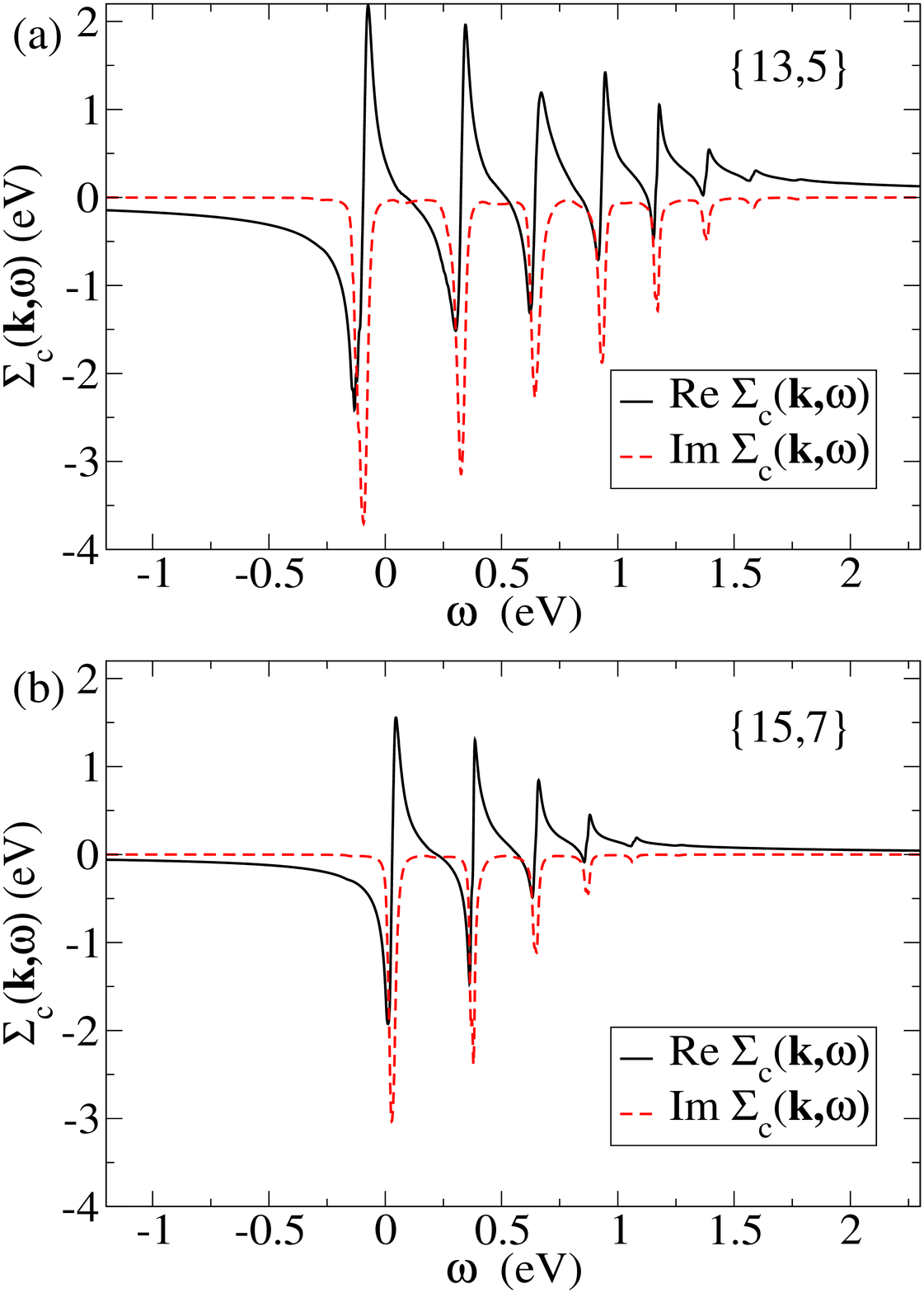,width=0.8\linewidth,clip=}
\caption{\label{SelfEnergy}(Color online) The real and imaginary parts of the 
SCBA conduction-band electron self-energy in the (a) $\{L=13,R=5\}$ and (b) $\{L=15,R=7\}$ 
graphene antidot lattices for $\mathbf{k}=0.2\:\mathbf{b}$, 
where $\mathbf{b}=(0,1)*4\pi/(3La)$.}
\end{figure}

\begin{figure}[t!]
\hspace*{-7.5mm}
\epsfig{file=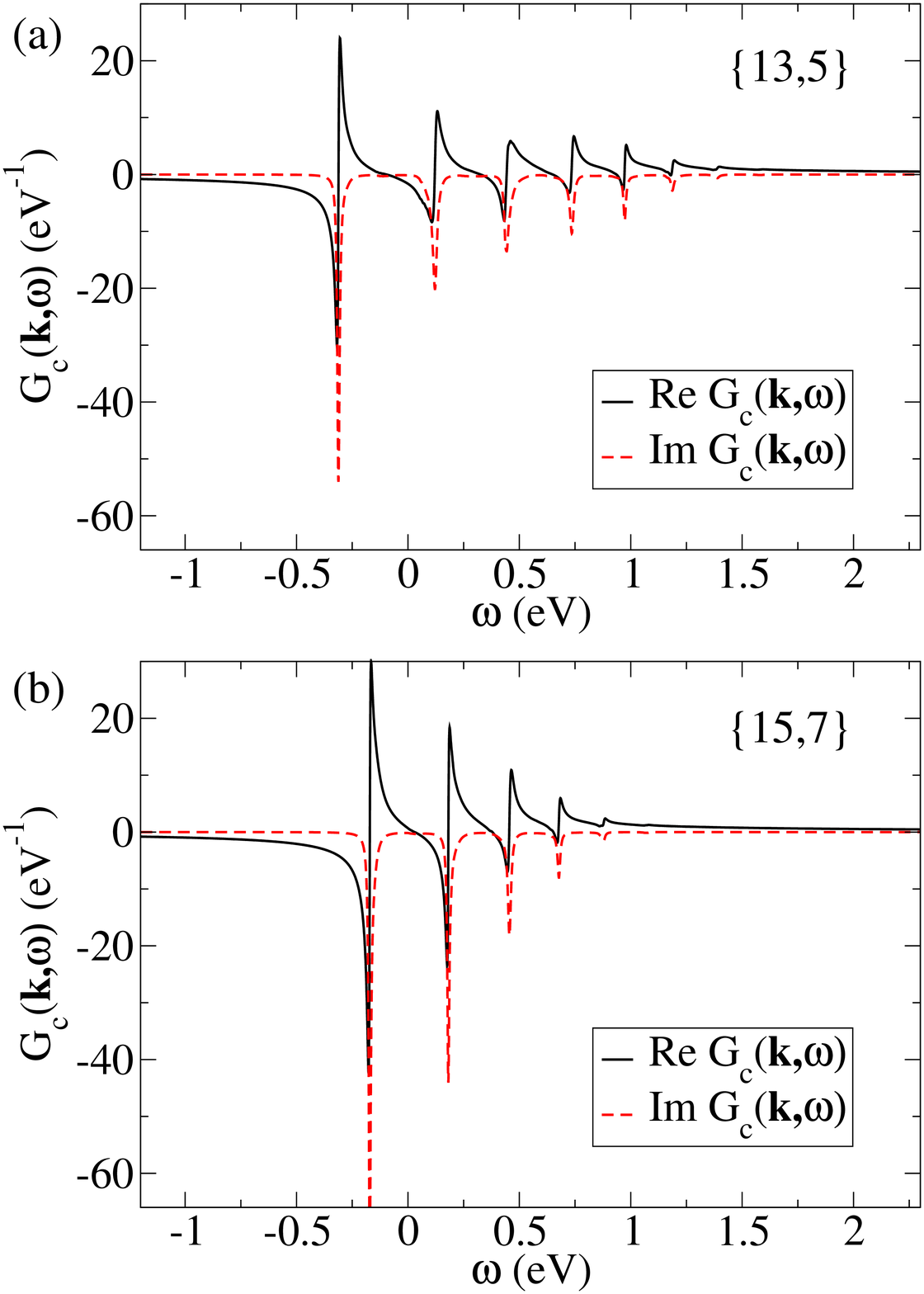,width=0.8\linewidth,clip=}
\caption{\label{GreenFunction}(Color online) The SCBA electron propagators 
in the (a) $\{L=13,R=5\}$ and (b) $\{L=15,R=7\}$ graphene antidot lattices 
for $\mathbf{k}=0.2\:\mathbf{b}$, where $\mathbf{b}=(0,1)*4\pi/(3La)$.}
\end{figure}

The last equation is solved self-consistently for electronic (conduction-band) self-energy 
$\Sigma_{\textrm{c}}(\mathbf{k},\omega)$ using the method of iteration, with the self-energy 
from Rayleigh-Schr\"{o}dinger perturbation theory~\cite{Vukmirovic+:10} 
taken as an initial guess in the iterative process at every 
$\mbf{k}$ and $\omega$. The resulting SCBA self-energies for the $\{L=13,R=5\}$ and $\{L=15,R=7\}$ 
antidot lattices are depicted in Fig.~\ref{SelfEnergy}(a) and (b), respectively. 
In this self-energy calculation, a broadening parameter $\delta$ was used, 
with a value of $5$\:meV that corresponds to the resolution 
of state-of-the-art photoemission experiments.~\cite{DamascelliARPES:04} 
The SCBA electron Green's function is then evaluated based on Eq.~\eqref{SCBAelprop},
with a typical example shown in Fig.~\ref{GreenFunction}. The obtained results are 
to be compared with future ARPES measurements of the single-particle spectral function 
$A_{\textrm{c}}(\mathbf{k},\omega)=-\Im G_{\textrm{c}}(\mathbf{k},\omega)/\pi$.

In the case of very weak e-ph interaction the peaks in the spectral function
appear at energies that are shifted from the bottom of the dressed-electron band by a characteristic
phonon energy. In the case at hand, the positions of peaks result from a complex interplay 
of multiple phonon modes involved and momentum-dependent e-ph coupling. The fact that the 
second peak is shifted from the first one by an energy larger than the maximal phonon energy 
in the system is an additional signature of strong e-ph interaction and polaronic behavior.

The renormalized conduction-band electron dispersion $E_{\textrm{c}}(\mathbf{k})$ 
is obtained by solving self-consistently the equation~\cite{Kirkegaard+:05} 
\begin{equation}\label{selfcon}
E_{\textrm{c}}(\mathbf{k})=\varepsilon_{\textrm{c}}(\mathbf{k})
+\Re\Sigma_{\textrm{c}}[\mathbf{k},E_{\textrm{c}}(\mathbf{k})] \:,
\end{equation}
where $\Re\Sigma_{\textrm{c}}(\mathbf{k},\omega)$ is the real part of the SCBA 
self-energy. The quasiparticle weight $Z_{\textrm{c}}(\mathbf{k})$ is then
computed for different nonzero quasimomenta $\mathbf{k}$ using the 
standard expression~\cite{MahanBook} 
\begin{equation}
Z_{c}(\mathbf{k})=\Big[1-\frac{\partial}{\partial\omega}\Re\Sigma_{c}
(\mathbf{k},\omega)\big|_{\omega=E_{c}(\mathbf{k})}\Big]^{-1} \:.
\end{equation}
The obtained values for the $\{L=9,R=5\}$ antidot lattice, for example, 
are in the range between $0.33$ and $0.37$, all of them being 
larger than $Z_{\textrm{c}}(\mathbf{k}=0)\approx 0.208$ obtained using 
Rayleigh-Schr\"{o}dinger perturbation theory. The increase of the spectral weight
away from $\mathbf{k}=0$ is intimately related to the fact that at larger 
total polaron momentum the momenta of phonons relevant in the scattering
process are also larger. Accordingly, the e-ph coupling that predominantly decreases
with phonon momentum (recall Sec.~\ref{PhononRenormA}) leads to a smaller renormalization.

In order to characterize the phonon-induced renormalization, it is of 
interest to determine the direction-dependent e-ph mass enhancement factor 
\begin{equation}
\lambda_{\textrm{me}}^{(\alpha)}\equiv
\left(\frac{m_{\textrm{eff}}}{m^{*}_{\textrm{e}}}
\right)_{\alpha}-1 \quad (\alpha=x,y)\:,
\end{equation}
\noindent where $m_{\textrm{eff}}$ is the effective (in the presence of phonons)
and $m^{*}_{\textrm{e}}$ the bare-band mass. Starting from the defining self-consistent 
relation for the renormalized dispersion [Eq.~\eqref{selfcon}], it is straightforward to show that 
(see Appendix~\ref{renormdirect} for a derivation consistent with our
present notation; alternatively, see Ref.~\onlinecite{MahanBook})
\begin{equation}\label{massrenormalpha} 
\left(\frac{m_{\textrm{eff}}}{m^{*}_{\textrm{e}}}\right)_{\alpha}
=\frac{\displaystyle 1-\frac{\partial}{\partial\omega}\Re\Sigma_{\textrm{c}}
(\mathbf{k},\omega)\big|_{\mathbf{k}=0,\omega=E_{\textrm{c}}(0)}}
{\displaystyle 1+\frac{\partial}{\partial\varepsilon_{\textrm{c}}
(\mathbf{k}_{\alpha})}\Re\Sigma_{\textrm{c}}(\mathbf{k}_{\alpha},\omega)
\big|_{\mathbf{k}_{\alpha}=0,\omega=E_{\textrm{c}}(0)}}  \:,
\end{equation}
\noindent where $\mathbf{k}_{\alpha}\equiv (\mathbf{k}\cdot\mathbf{e}_{\alpha})
\mathbf{e}_{\alpha}\equiv k_{\alpha}\mathbf{e}_{\alpha}$. It then directly follows that
\begin{equation}\label{lambdamealpha}
\lambda_{\textrm{me}}^{(\alpha)}=\frac{Z_{\textrm{c}}^{-1}(0)}
{\displaystyle 1+\frac{\partial}{\partial\varepsilon_{\textrm{c}}
(\mathbf{k}_{\alpha})}\Re\Sigma_{\textrm{c}}(\mathbf{k}_{\alpha},\omega)
\big|_{\mathbf{k}_{\alpha}=0,\omega=E_{\textrm{c}}(0)}}-1\:.
\end{equation}
The last quantity can be obtained by a numerical differentiation of 
$\Re\Sigma_{\textrm{c}}$, using previously obtained values for $Z_{\textrm{c}}^{-1}(0)$. 
For the $\{L=9,R=5\}$ lattice, our calculations yield the values $\lambda_{\textrm{me}}^{(x)}=2.411$ 
and $\lambda_{\textrm{me}}^{(y)}=2.448$ [the calculated values of the respective denominators of the 
fraction in Eq.~\eqref{lambdamealpha} are 1.4106 and 1.3954]. For the $\{L=13,R=5\}$ lattice we 
obtain $\lambda_{\textrm{me}}^{(x)}=2.389$ and $\lambda_{\textrm{me}}^{(y)}=2.081$, respectively.
In the $\{L=15,R=7\}$ lattice case, we get $\lambda_{\textrm{me}}^{(x)}=1.365$ and 
$\lambda_{\textrm{me}}^{(y)}=1.255$.

The obtained numbers suggest that the anisotropy in the e-ph mass enhancement 
is relatively weakly pronounced, despite the existence of anisotropy in the momentum-dependent e-ph 
coupling [recall Fig.~\ref{VertexFunc}]. This can be qualitatively explained by observing 
that the latter anisotropy of the e-ph coupling is smallest for the highest-energy 
phonon modes which yield the strongest coupling [for an illustration, note the variation of the values of 
$|\gamma^{\lambda}_{\textrm{cc}}(\mbf{k}=0,\mbf{q})|$ in Fig.~\ref{VertexFunc}(a) and (c), 
as opposed to those in (b) and (d)]. We can therefore draw the conclusion that the 
overall effective-mass anisotropy in graphene antidot lattices is predominantly determined by 
the anisotropy of the bare band mass itself, rather than that stemming from phonon-related effects. 

\section{Discussion} \label{discuss}
The strong renormalization obtained suggests that the charge carriers in the system 
acquire polaronic character. Indeed, it is not surprising to have polaronic charge 
carriers in a narrow-band system. This is a commonplace, for example, in organic semiconductors, 
such as organic molecular crystals, where narrow electronic bands result from 
weak van der Waals intermolecular bonding and the e-ph interaction has similar 
(slightly nonlocal, but still short-ranged) character.~\cite{Hannewald+Hatch+Koller} 
Compared to these systems, graphene antidot lattices have yet narrower conduction bands 
(e.g., bare conduction-electron band widths in the polyacene organic crystals can be as 
large as $700$\:meV) and lower dimensionality, both of these factors being conducive 
to the larger mass enhancement found here. 

The strong-coupling regimes in interacting e-ph systems are usually explored 
using exact diagonalization or variational methods. The criteria for the onset
of a (smooth) polaron crossover with the increasing e-ph coupling strength 
are typically based either on the behavior of the average number of phonons in 
the polaron ground state -- which in the crossover region grows precipitously -- or on the 
behavior of the quasiparticle spectral weight, which in the same regime 
undergoes a sharp decrease. For a system with short-range, non-polar e-ph interaction
(regardless of its concrete form), as is generally the case in covalently-bonded materials such 
as the one considered here, the polaron crossover has the nature 
of a change from a quasi-free electron to a small polaron (rather than being a 
large-to-small polaron crossover, for which a long-range character of the e-ph interaction is a 
crucial ingredient). However, for a system of such complexity as studied 
here the use of the aforementioned methods would clearly be inconceivable even with a 
single ``effective'' phonon mode, thus the criteria for polaronic behavior cannot be 
utilized in a strict sense. The perturbative approach is the most powerful technique at our disposal.  
Obviously, the obtained results are not expected to be valid quantitatively. However, they should 
still be qualitatively correct, providing an unambiguous evidence for the relevance of phonons in 
graphene antidot lattices. In addition, it should be stressed that the extent to which 
the obtained results can be improved is limited: while the accuracy in obtaining the phonon spectra 
can certainly be improved, due to the prohibitive system size the band structure remains tractable only at 
the level of the tight-binding method. Finally, even if DFT calculations on such a complex
system were feasible, the validity of DFT when calculating e-ph coupling in two-dimensional 
systems has been called into question.~\cite{Lazzeri+:08}

An experimental verification of the polaronic nature of carriers in our system should be
possible in future optical absorption measurements. Some qualitative predictions can already 
be made based on general features of (intraband) optical absorption in systems with Peierls-type 
(SSH) e-ph coupling.~\cite{Capone+:97} In the polaronic regime, the ground state of such systems is 
characterized by a shortened bond -- being a result of two adjacent sites shifting towards one 
another -- on which the electron is localized. The ground state is then the even combination of the 
two local electron states. Viewed through simple Franck-Condon-type picture, the optical absorption in 
such systems proceeds through two different channels: one in which the electron is excited onto another bond 
(which is not shortened), and the other one in which the electron is excited from the (even-symmetry) 
ground state to the local odd-symmetry state on the shortened bond itself. The excitation energy in the 
first process corresponds to the energy gained by the lattice distortion (compared to the undistorted one) 
and is given by $2E_{\rm p}$, where $E_{\rm p}$ is the polaron binding energy. In the second absorption
channel, which corresponds to a transition between a state where the electron energy is lower by $2E_{\rm p}$
with respect to the undistorted lattice to a state where it is higher by the same amount, the excitation energy
is twice as large ($4E_{\rm p}$). When the finiteness of the relevant phonon frequency is taken into account,
the absorption peaks at energies $2E_{\rm p}$ and $4E_{\rm p}$ from the Franck-Condon picture 
broaden into bands characterized by phonon features separated by the characteristic phonon energy. 

An immediate implication of our results is that the charge transport in graphene antidot 
lattices -- at variance with that of their ``parent material'' --  ought to be modelled by 
taking into account the inelastic degrees of freedom. This was done previously with carbon 
nanotubes,~\cite{Roche+:05} where the inelastic effects in transport are relevant only in 
the high-bias regime as a consequence of the carrier energy exceeding that of an optical phonon 
(with energy of $\approx 200$\:meV).~\cite{CarbonElectronics} 
However, in our system there are also optical phonons at far lower energies. 
Thus the inelastic effects should be even more pronounced than in carbon nanotubes
and not necessarily restricted to the high-bias regime.
  
It is worthwhile noting that the recently fabricated graphene antidot lattices~\cite{Bai+:10,Kim+:10} 
comprise even larger number of atoms per unit cell ($N_{\textrm{at}}\gtrsim 5,000$) than
the ones discussed here. For comparison, their periods are in the range $27-39$\:nm,
while the largest period of the superlattices studied here is around $8.2$\:nm. 
Thus it is clearly of interest to extend the present study to such cases.
However, a detailed treatment of the e-ph coupling effects, whose
complexity scales roughly with $N^{3}_{\textrm{at}}$, for systems of such size would be extremely 
computationally demanding even with the methodology employed in the present work. Such cases can, however,
be addressed within the effective-mass approximation for electronic band structure
(wherein the wave functions are expressed through slowly-varying envelope functions) and with a
continuum treatment of phonons (deformation potential), an approach well established in other 
systems.~\cite{Ando++nanotube} Knowing that these large-period lattices have larger conduction-band
widths (in addition to having smaller band gaps) it is likely that the e-ph coupling effects 
will be quantitatively less important than in the lattices studied here.

Another important issue for future investigation is the interplay of (short-range) 
Peierls-type coupling in graphene antidot lattices with the long-range (Fr\"{o}hlich-type) 
e-ph coupling at their interface with polarizable substrates such as SiC or SiO$_{2}$.~\cite{Fratini:08}
This interplay between the two e-ph coupling mechanisms is likely to be relevant in a 
field-effect-transistor geometry, where a graphene antidot lattice
can be used as the semiconducting channel and SiO$_{2}$ as a high-$\kappa$ gate dielectric.~\cite{Bai+:10} 
Quite recently, motivated by its relevance in organic field-effect transistors,
such interplay has been investigated within a simplified one-dimensional model.~\cite{DeFilippis+:10}
The main conclusion of this study was that even a rather weak SSH coupling
is sufficient to obtain a polaronic behavior in the simultaneous presence 
of a moderate polar coupling. In graphene antidot lattices, as a consequence of
their different geometry and the ensuing momentum dependence of the e-ph coupling 
(compared to the genuine SSH coupling; recall discussion in Sec.~\ref{PhononRenormA}) 
such issues would need to be reconsidered. 
 
\section{Summary and Conclusions}\label{sumconc}

To summarize, we have studied the effects of electron-phonon interaction
in graphene antidot lattices. Using realistic band-structure and phonon-spectra 
calculations as an input, we have described the system based on a model that
accounts for the phonon-modulation of electronic hopping integrals -- Peierls' type
electron-phonon coupling. We have demonstrated a significant electron-phonon mass enhancement 
in this system, which is likely to be verified in future ARPES measurements.

The main message of the present work is that in graphene antidot lattices -- in contrast to
graphene, their parent material -- optical phonons, particularly the highest-energy ones, 
play a very important role. This can be attributed to the narrow-band character of these systems, 
and, especially, to the peculiar momentum dependence of the relevant electron-phonon interaction 
which is strongest for small phonon momenta.

Apart from being related to a particular family of graphene-based nanostructured systems  
of keen current interest,~\cite{Bai+:10,Kim+:10} our study also bears some fundamental importance. 
Namely, it provides an insight into the relevance and physical consequences of strongly
momentum-dependent electron-phonon interactions. The body of work concerned with such interactions is 
fairly limited, even at the level of model investigations,~\cite{BerciuMomDependent} 
in spite of their importance in realistic electronic materials.

The present work bodes well for further studies of graphene antidot lattices, in the 
finite-temperature case and in the presence of a substrate and/or metallic 
contacts,~\cite{GrapheneMetalCont} all of these aspects being of particular 
relevance for future room-temperature, carbon-based electronic devices. 
By providing evidence for the relevance of the inelastic degrees of freedom, our findings 
are expected to also facilitate future studies of electronic transport in these systems. 
Detailed investigations of electron-phonon-interaction effects in other emerging classes of 
graphene-based superlattices~\cite{GrapheneSuperlattPRB} are clearly called for.

\begin{acknowledgments}
The authors are grateful to M. Vanevi\'c for useful comments.
V.M.S. acknowledges interesting discussions with J.C. Egues, 
D.C. Glattli, and B. Trauzettel. This work was financially 
supported by the Swiss NSF and the NCCR Nanoscience.
\end{acknowledgments}
\appendix
\section{Derivation of the electron-phonon-coupling vertex functions \label{vertfuncapp}}
By introducing the matrix $\mathbf{C}_{\veck}$ (for each $\mathbf{k}$ in the BZ) 
such that $C_{m}^{n,\veck}=(\mathbf{C}_{\veck})_{nm}$, 
Eq.~\eqref{BlochWF} can be rewritten as
\begin{equation} \label{}
\bm{\gpsi}_{\mbf k}(\mbf r)=
\mathbf{C}_{\veck}\bm{\gphi}_{\veck}(\vecr) \:,
\end{equation}
\noindent where $\bm{\gpsi}_{\mbf k}(\mbf r)\equiv[\gpsi_{n\mbf k}(\mbf r)|\:n=1,
\ldots,N_{\textrm{at}}]^{T}$ and $\bm{\gphi}_{\veck}(\vecr)\equiv[\phi_{m\veck}
(\mbf r)|\:m=1,\ldots,N_{\textrm{at}}]^{T}$ are $N_{\textrm{at}}$-component vectors. 
The inverse of the last equation 
\begin{equation} \label{}
\bm{\gphi}_{\mbf k}(\mbf r)=
\mathbf{C}^{-1}_{\veck}\bm{\gpsi}_{\veck}(\vecr) \:,
\end{equation}
\noindent when recast componentwise reads
\begin{equation} \label{ddeq}
\gphi_{m\veck}(\vecr)=\sum_{n} 
(\mathbf{C}^{-1}_{\veck})_{mn}\gpsi_{n\mbf k}(\mbf r) \:.
\end{equation}
\noindent Given that $C_{m}^{n,\veck}$ 
are the coefficients of a linear transformation [recall Eq.~\eqref{BlochWF}] 
between two complete orthonormal sets of functions, 
the matrix $\mathbf{C}_{\veck}$ is unitary, so that
$(\mathbf{C}^{-1}_{\veck})_{mn}=(\mathbf{C}^{\dagger}_{\veck})_{mn}
=(C_{m}^{n,\veck})^{*}$. Consequently, Eq.~\eqref{ddeq} is equivalent to
\begin{equation} \label{druga}
\gphi_{m\veck}(\vecr)=\sum_{n}(C_{m}^{n,\veck})^{*}
\gpsi_{n\mbf k}(\mbf r) \:,
\end{equation}
\noindent and, combined with the inverse Fourier 
transformation 
\begin{equation} \label{invft}
\gvphi(\vecr-\mbf R -\mbf d_m)=\frac{1}{\sqrt{N}}
\sum_{\mbf k}\gphi_{m\veck}(\vecr)\:e^{-i\veck \cdot \mbf R} \:,
\end{equation}
\noindent readily implies that
\begin{equation}
\hat{a}^\dg_{\mathbf{R}+\mathbf{d}_{m}} =
\frac{1}{\sqrt{N}}\sum_{n,\mbf k}(C_{m}^{n,\veck})^{*}
e^{-i\veck\cdot\mbf R}\hat{a}^\dg_{n\mbf k} \:.
\end{equation}
\noindent If, in addition, we assume that for a given vector $\mathbf{d}_{m}$,
the vector $\mathbf{d}_{m}+\bm{\delta}$ denotes a site within the same unit cell
(i.e., hopping takes place between two lattice sites within the same unit cell), 
it also holds that
\begin{equation} \label{delneb}
\hat{a}_{\mathbf{R}+\mathbf{d}_{m}+\bm{\delta}} =
\frac{1}{\sqrt{N}}\sum_{n,\mbf k}C_{m+\delta}^{n,\veck}
\:e^{i\veck\cdot\mbf R}\hat{a}_{n\mbf k}\:.
\end{equation}
From the last two equations, combined with the expression for phonon displacements
\begin{equation}\label{phonondisp}
\hat{\mathbf{u}}_{\lambda,\mathbf{R}+\mathbf{d}_{m}}\equiv
\frac{1}{\sqrt{N}}\sum_{\mathbf{q}}\frac{e^{i\mathbf{q}\cdot\mathbf{R}}
(\hat{b}^\dg_{-\mathbf{q},\lambda}+\hat{b}_{\mathbf{q},\lambda})}
{\sqrt{2M\omega_{\lambda}(\mathbf{q})}}\:\mathbf{v}^{\lambda}_{m}(\mathbf{q}) \:,
\end{equation}
we readily obtain
\begin{multline} \label{}
\sum_{\mathbf{R}}(\hat{a}^\dg_{\mathbf{R}+\mathbf{d}_{m}+\bm{\delta}}
\hat{a}_{\mathbf{R}+\mathbf{d}_{m}}+\hc)
\big[\hat{\mathbf{u}}_{\lambda,\mathbf{R}+\mathbf{d}_{m}+\bm{\delta}}
-\hat{\mathbf{u}}_{\lambda, \mathbf{R}+\mathbf{d}_{m}}\big]\cdot 
\bar{\bm{\delta}}\\
=\frac{1}{\sqrt{N}}\sum_{n,n',\mbf k,\mbf q}\frac{1}{{\sqrt{2M\omega_\lambda(\mbf q)}}}
\:\bar{\bm{\delta}}\cdot [\mbf v^\lambda_{m+\delta}
(\mbf q)-\mbf v^\lambda_m(\mbf q)]\\
\times\big[(C^{n, \mbf k+\mbf q}_{m+\delta})^* C^{n',\mbf k}_m+ 
(C^{n, \mbf k + \mbf q}_m)^* C^{n', \mbf k}_{m+\delta}\big]\\
\times\hat{a}_{n,\mathbf{k+q}}^{\dagger}\hat{a}_{n',\mathbf{k}}
(\hat{b}_{-\mathbf{q},\lambda}^{\dagger}+\hat{b}_{\mathbf{q},\lambda}) \:.
\end{multline}
\noindent The last expression straightforwardly leads to the
part of the Hamiltonian in Eq.~\eqref{mscoupling} that corresponds
to the hopping within the same unit cell, from which we can read out 
the expression for $V_{nn'}^\lambda (\mbf k ,\mbf q)$ in Eq.~\eqref{vertex1}. 

If, on the other hand, for a vector $\mathbf{d}_{m}$ in the given unit cell, 
the nearest-neighbor vector $\mathbf{d}_{m}+\bm{\delta}$ belongs to one of the 
adjacent unit cells -- displaced from the given one by the vector $\mbf a$ 
[one of the vectors $\pm \mbf a_1$, $\pm \mbf a_2$, 
$\pm(\mbf a_1-\mbf a_2)$] -- Eq.~\eqref{delneb} ought to be replaced by
\begin{equation} \label{delnebout}
\hat{a}_{\mathbf{R}+\mathbf{d}_{m}+\bm{\delta}} =
\frac{1}{\sqrt{N}}\sum_{n,\mbf k}C_{m_{1}}^{n,\veck}
e^{i\veck\cdot(\mbf R+\mbf a)}\hat{a}_{n\mbf k}\:,
\end{equation}
\noindent where in the last expression $m_{1}=m_{1}(\bm{\delta})$ 
($m_{1}\in \{1,\ldots,N_{\textrm{at}}\}$) 
is chosen in such a way as to satisfy the condition 
$\mbf d_m + \bm \delta =\mbf a + \mbf d_{m_1}$. In an analogous manner,
the expression for $\hat{\mathbf{u}}_{\lambda,\mathbf{R}
+\mathbf{d}_{m}+\bm{\delta}}=\hat{\mathbf{u}}_{\lambda,\mathbf{R}
+\mbf a + \mbf d_{m_1}}$ should then read [cf. Eq.~\eqref{phonondisp}]
\begin{equation} \label{phondisout}
\hat{\mathbf{u}}_{\lambda,\mathbf{R}+\mathbf{d}_{m}+
\bm{\delta}}\equiv\frac{1}{\sqrt{N}}\sum_{\mathbf{q}}
\frac{e^{i\mathbf{q}\cdot(\mathbf{R}+\mathbf{a})}
(\hat{b}^\dg_{-\mathbf{q},\lambda}+\hat{b}_{\mathbf{q},\lambda})}
{\sqrt{2M\omega_{\lambda}(\mathbf{q})}}\:
\mathbf{v}^{\lambda}_{m_{1}}(\mathbf{q}) \:.
\end{equation}
\noindent Now, by repeating the steps in the above derivation of the 
expression for $V_{nn'}^\lambda (\mbf k ,\mbf q)$, with regard for 
Eqs.~\eqref{delnebout} and \eqref{phondisout}, we
easily recover expression \eqref{vertex2} for 
$W_{nn'}^\lambda (\mbf k ,\mbf q)$.

\section{\label{renormdirect}Derivation of Eq.~(\ref{massrenormalpha})}

We start from the defining self-consistent relation~\cite{Kirkegaard+:05} 
for the renormalized conduction-band electron dispersion:
\begin{equation}\label{ }
 E_{\textrm{c}}(\mathbf{k})=\varepsilon_{\textrm{c}}(\mathbf{k})
 +\Re\Sigma_{\textrm{c}}
 [\mathbf{k},E_{\textrm{c}}(\mathbf{k})] \:.
\end{equation}
\noindent By expanding the second term on the right-hand side of the last 
equation around $\mathbf{k}=0$, we readily obtain 
\begin{multline}\label{ }
E_{\textrm{c}}(\mathbf{k})=\varepsilon_{\textrm{c}}(\mathbf{k})
+\mathbf{k}\cdot\frac{\partial}{\partial\mathbf{k}}\Re\Sigma_{\textrm{c}}
(\mathbf{k},\omega)\big|_{\mathbf{k}=0,\omega=E_{\textrm{c}}(0)}\\
+E_{\textrm{c}}(\mathbf{k})\frac{\partial}{\partial\omega}\Re\Sigma_{\textrm{c}}
(\mathbf{k},\omega)\big|_{\mathbf{k}=0,\omega=E_{\textrm{c}}(0)}
+\mathcal{O}(\mathbf{k}^{2})\:,
\end{multline}
\noindent which combined with the identity
\begin{equation}\label{ }
 E_{\textrm{c}}(0)=\varepsilon_{\textrm{c}}(0)
 +\Re\Sigma_{\textrm{c}}[0,E_{\textrm{c}}(0)] 
\end{equation}
\noindent leads to the relation
\begin{multline}\label{longeq}
\big[E_{\textrm{c}}(\mathbf{k})-E_{\textrm{c}}(0)\big]
\left[1-\frac{\partial}{\partial\omega}\Re\Sigma_{\textrm{c}}
(\mathbf{k},\omega)\big|_{\mathbf{k}=0,\omega=E_{\textrm{c}}(0)}\right] \\
=\varepsilon_{\textrm{c}}(\mathbf{k})-\varepsilon_{\textrm{c}}(0)
+\mathbf{k}\cdot\frac{\partial}{\partial\mathbf{k}}\Re
\Sigma_{\textrm{c}}(\mathbf{k},\omega)\big|_{\mathbf{k}=0,\omega=E_{\textrm{c}}(0)}\\
+\mathcal{O}(\mathbf{k}^{2})\:.
\end{multline}
The mass renormalization in direction $\alpha$ is given by
\begin{equation}\label{ }
\left(\frac{m_{\textrm{eff}}}{m^{*}_{\textrm{e}}}\right)_{\alpha}
=\lim_{\mathbf{k}_{\alpha}\rightarrow 0}\frac{\varepsilon_{\textrm{c}}
(\mathbf{k}_{\alpha})-\varepsilon_{\textrm{c}}(0)}{E_{\textrm{c}}
(\mathbf{k}_{\alpha})-E_{\textrm{c}}(0)} \:.
\end{equation}
\noindent where $\mathbf{k}_{\alpha}\equiv (\mathbf{k}\cdot\mathbf{e}_{\alpha})
\mathbf{e}_{\alpha}\equiv k_{\alpha}\mathbf{e}_{\alpha}$. Using the special case 
$\mathbf{k}=\mathbf{k}_{\alpha}$ of Eq.~\eqref{longeq}, after some straightforward 
manipulations, we finally obtain
\begin{equation}
\left(\frac{m_{\textrm{eff}}}{m^{*}_{\textrm{e}}}\right)_{\alpha}
=\frac{\displaystyle 1-\frac{\partial}{\partial\omega}\Re\Sigma_{\textrm{c}}
(\mathbf{k},\omega)\big|_{\mathbf{k}=0,\omega=E_{\textrm{c}}(0)}}
{\displaystyle 1+\frac{\partial}{\partial\varepsilon_{\textrm{c}}
(\mathbf{k}_{\alpha})}\Re\Sigma_{\textrm{c}}(\mathbf{k}_{\alpha},\omega)
\big|_{\mathbf{k}_{\alpha}=0,\omega=E_{\textrm{c}}(0)}}  \:.
\end{equation}

\bibliography{Graphene,polaronref}

\begin{thebibliography}{56}
\expandafter\ifx\csname natexlab\endcsname\relax\def\natexlab#1{#1}\fi
\expandafter\ifx\csname bibnamefont\endcsname\relax
  \def\bibnamefont#1{#1}\fi
\expandafter\ifx\csname bibfnamefont\endcsname\relax
  \def\bibfnamefont#1{#1}\fi
\expandafter\ifx\csname citenamefont\endcsname\relax
  \def\citenamefont#1{#1}\fi
\expandafter\ifx\csname url\endcsname\relax
  \def\url#1{\texttt{#1}}\fi
\expandafter\ifx\csname urlprefix\endcsname\relax\def\urlprefix{URL }\fi
\providecommand{\bibinfo}[2]{#2}
\providecommand{\eprint}[2][]{\url{#2}}

\bibitem[{Gra({\natexlab{a}})}]{GrapheneReviews}
\bibinfo{note}{For a review, see A. H. Castro Neto, F. Guinea, N. M. R. Peres,
  K. S. Novoselov, and A. K. Geim, Rev. Mod. Phys. ${\mathbf{81}}$, 109 (2009);
  A. K. Geim, Science ${\mathbf{324}}$, 1530 (2009).}

\bibitem[{Moi()}]{MoireSuperlattice}
\bibinfo{note}{J. M. B. Lopes dos Santos, N. M. R. Peres, and A. H. Castro
  Neto, Phys. Rev. Lett. $\mathbf{99}$, $256802$ ($2007$).}

\bibitem[{Ber({\natexlab{a}})}]{BerkeleySuperlatt}
\bibinfo{note}{C.-H. Park, L. Yang, Y.-W. Son, M. L. Cohen, and S. G. Louie,
  Nature Phys. ${\mathbf{4}}$, $213$ ($2008$).}

\bibitem[{\citenamefont{Pedersen et~al.}(2008)\citenamefont{Pedersen, Flindt,
  Pedersen, Mortensen, Jauho, and Pedersen}}]{PedersenAntidot:08}
\bibinfo{author}{\bibfnamefont{T.~G.} \bibnamefont{Pedersen}},
  \bibinfo{author}{\bibfnamefont{C.}~\bibnamefont{Flindt}},
  \bibinfo{author}{\bibfnamefont{J.}~\bibnamefont{Pedersen}},
  \bibinfo{author}{\bibfnamefont{N.~A.} \bibnamefont{Mortensen}},
  \bibinfo{author}{\bibfnamefont{A.-P.} \bibnamefont{Jauho}}, \bibnamefont{and}
  \bibinfo{author}{\bibfnamefont{K.}~\bibnamefont{Pedersen}},
  \bibinfo{journal}{Phys. Rev. Lett.} \textbf{\bibinfo{volume}{100}},
  \bibinfo{eid}{136804} (\bibinfo{year}{2008}).

\bibitem[{Gra({\natexlab{b}})}]{GrapheneSuperlattPRB}
\bibinfo{note}{A. Isacsson, L. M. Jonsson, J. M. Kinaret, and M. Jonson, Phys.
  Rev. B ${\mathbf{77}}$, $035423$ ($2008$); H. Sevin{\c{c}}li, M. Topsakal,
  and S. Ciraci, {\em ibid.} ${\mathbf{78}}$, $245402$ ($2008$); Y. P. Bliokh,
  V. Freilikher, S. Savel'ev, and F. Nori, {\em ibid.} ${\mathbf{79}}$,
  $075123$ ($2009$); L.-G. Wang and S.-Y. Zhu, {\em ibid.} ${\mathbf{81}}$,
  $205444$ ($2010$).}

\bibitem[{Bal()}]{Balog+:10}
\bibinfo{note}{R. Balog, B. J\o rgensen, L. Nilsson, M. Andersen, E. Rienks, M.
  Bianchi, M. Fanetti, E. L\ae gsgaard, A. Baraldi, S. Lizzit, Z. Sljivancanin,
  F. Besenbacher, B. Hammer, T. G. Pedersen, P. Hofmann, and L. Hornek\ae r,
  Nat. Mater. ${\mathbf{9}}$, $315$ ($2010$).}

\bibitem[{\citenamefont{Shima and Aoki}(1993)}]{Shima+Aoki:93}
\bibinfo{author}{\bibfnamefont{N.}~\bibnamefont{Shima}} \bibnamefont{and}
  \bibinfo{author}{\bibfnamefont{H.}~\bibnamefont{Aoki}},
  \bibinfo{journal}{Phys. Rev. Lett.} \textbf{\bibinfo{volume}{71}},
  \bibinfo{pages}{4389} (\bibinfo{year}{1993}).

\bibitem[{\citenamefont{Bai et~al.}(2010)\citenamefont{Bai, Zhong, Jiang,
  Huang, and Duan}}]{Bai+:10}
\bibinfo{author}{\bibfnamefont{J.}~\bibnamefont{Bai}},
  \bibinfo{author}{\bibfnamefont{X.}~\bibnamefont{Zhong}},
  \bibinfo{author}{\bibfnamefont{S.}~\bibnamefont{Jiang}},
  \bibinfo{author}{\bibfnamefont{Y.}~\bibnamefont{Huang}}, \bibnamefont{and}
  \bibinfo{author}{\bibfnamefont{X.}~\bibnamefont{Duan}},
  \bibinfo{journal}{Nat. Nanotechnol.} \textbf{\bibinfo{volume}{5}},
  \bibinfo{pages}{190} (\bibinfo{year}{2010}).

\bibitem[{Sem()}]{SemiconAntidot}
\bibinfo{note}{K. Ensslin and P. M. Petroff, Phys. Rev. B $\mathbf{41}$, 12307
  (1990); D. Weiss {\em et al.}, Phys. Rev. Lett. $\mathbf{66}$, 2790 (1991).}

\bibitem[{Ant()}]{AntidotExper}
\bibinfo{note}{T. Shen, Y. Q. Wu, M. A. Capano, L. R. Rokhinson, L. W. Engel,
  and P. D. Ye, Appl. Phys. Lett. ${\mathbf{93}}$, 122102 (2008).}

\bibitem[{\citenamefont{Vanevi\'{c} et~al.}(2009)\citenamefont{Vanevi\'{c},
  Stojanovi\'{c}, and Kindermann}}]{Vanevic+:09}
\bibinfo{author}{\bibfnamefont{M.}~\bibnamefont{Vanevi\'{c}}},
  \bibinfo{author}{\bibfnamefont{V.~M.} \bibnamefont{Stojanovi\'{c}}},
  \bibnamefont{and}
  \bibinfo{author}{\bibfnamefont{M.}~\bibnamefont{Kindermann}},
  \bibinfo{journal}{Phys. Rev. B} \textbf{\bibinfo{volume}{80}},
  \bibinfo{pages}{045410} (\bibinfo{year}{2009}).

\bibitem[{Fue()}]{Fuerst++:09}
\bibinfo{note}{J. A. F{\"{u}}rst, J. G. Pedersen, C. Flindt, N. A. Mortensen,
  M. Brandbyge, T. G. Pedersen, and A.-P. Jauho, New J. Phys. ${\mathbf{11}}$,
  095020 (2009).}

\bibitem[{Ped()}]{PedersenOptical}
\bibinfo{note}{T. G. Pedersen, C. Flindt, J. Pedersen, A.-P. Jauho, N. A.
  Mortensen, and K. Pedersen, Phys. Rev. B ${\mathbf{77}}$, $245431$ ($2008$);
  R. Petersen and T. G. Pedersen, {\em ibid.} ${\mathbf{80}}$, $113404$
  ($2009$).}

\bibitem[{Kin()}]{KineziAntidot}
\bibinfo{note}{D. Yu, E. M. Lupton, M. Liu, W. Liu, and F. Liu, Nano Res.
  ${\mathbf{1}}$, 56 (2008); W. Liu, Z. F. Wang, Q. W. Shi, J. Yang, and F.
  Liu, Phys. Rev. B ${\mathbf{80}}$, 233405 (2009).}

\bibitem[{\citenamefont{Tworzyd{\l}o et~al.}(2006)\citenamefont{Tworzyd{\l}o,
  Trauzettel, Titov, Rycerz, and Beenakker}}]{Tworzydlo:06}
\bibinfo{author}{\bibfnamefont{J.}~\bibnamefont{Tworzyd{\l}o}},
  \bibinfo{author}{\bibfnamefont{B.}~\bibnamefont{Trauzettel}},
  \bibinfo{author}{\bibfnamefont{M.}~\bibnamefont{Titov}},
  \bibinfo{author}{\bibfnamefont{A.}~\bibnamefont{Rycerz}}, \bibnamefont{and}
  \bibinfo{author}{\bibfnamefont{C.~W.~J.} \bibnamefont{Beenakker}},
  \bibinfo{journal}{Phys. Rev. Lett.} \textbf{\bibinfo{volume}{96}},
  \bibinfo{pages}{246802} (\bibinfo{year}{2006}).

\bibitem[{Car()}]{CarbonElectronics}
\bibinfo{note}{See, e.g., P. Avouris, Z. Chen, and V. Perebeinos, Nat.
  Nanotechnol. $\mathbf{2}$, 605 (2007).}

\bibitem[{\citenamefont{M{\"{u}}ller et~al.}(2009)\citenamefont{M{\"{u}}ller,
  Br{\"{a}}uninger, and Trauzettel}}]{Mueller+:09}
\bibinfo{author}{\bibfnamefont{M.}~\bibnamefont{M{\"{u}}ller}},
  \bibinfo{author}{\bibfnamefont{M.}~\bibnamefont{Br{\"{a}}uninger}},
  \bibnamefont{and}
  \bibinfo{author}{\bibfnamefont{B.}~\bibnamefont{Trauzettel}},
  \bibinfo{journal}{Phys. Rev. Lett.} \textbf{\bibinfo{volume}{103}},
  \bibinfo{eid}{196801} (\bibinfo{year}{2009}).

\bibitem[{Kle()}]{KleinTunneling}
\bibinfo{note}{V. V. Cheianov and V. I. Fal'ko, Phys. Rev. B ${\mathbf{74}}$,
  $041403$ (2006).}

\bibitem[{\citenamefont{Kim et~al.}(2010)\citenamefont{Kim, Safron, Han,
  Arnold, and Gopalan}}]{Kim+:10}
\bibinfo{author}{\bibfnamefont{M.}~\bibnamefont{Kim}},
  \bibinfo{author}{\bibfnamefont{N.~S.} \bibnamefont{Safron}},
  \bibinfo{author}{\bibfnamefont{E.}~\bibnamefont{Han}},
  \bibinfo{author}{\bibfnamefont{M.~S.} \bibnamefont{Arnold}},
  \bibnamefont{and} \bibinfo{author}{\bibfnamefont{P.}~\bibnamefont{Gopalan}},
  \bibinfo{journal}{Nano Lett.} \textbf{\bibinfo{volume}{10}},
  \bibinfo{pages}{1125} (\bibinfo{year}{2010}).

\bibitem[{EPH()}]{EPHgeneralMarsiglio}
\bibinfo{note}{See, e.g., F. Marsiglio and J. P. Carbotte, in {\em The Physics
  of Conventional and Unconventional Superconductors}, edited by K. H.
  Bennemann and J. B. Ketterson (Springer-Verlag, Berlin, 2003), pp. 233--345.}

\bibitem[{SSH({\natexlab{a}})}]{SSH}
\bibinfo{note}{W. P. Su, J. R. Schrieffer, and A. J. Heeger, Phys. Rev. Lett.
  ${\mathbf {42}}$, $1698$ ($1979$); Phys. Rev. B ${\mathbf{22}}$, $2099$
  ($1980$).}

\bibitem[{Mah()}]{MahanEtAl}
\bibinfo{note}{L. M. Woods and G. D. Mahan, Phys. Rev. B ${\mathbf{61}}$,
  $10651$ (2000); G. D. Mahan, {\em ibid.} ${\mathbf{68}}$, $125409$ (2003).}

\bibitem[{Non()}]{NonlocalCouplingLong}
\bibinfo{note}{V. M. Stojanovi\'{c}, P. A. Bobbert, and M. A. J. Michels, Phys.
  Rev. B ${\mathbf{69}}$, $144302$ ($2004$); C. A. Perroni, E. Piegari, M.
  Capone, and V. Cataudella, {\em ibid.} ${\mathbf{69}}$, $174301$ ($2004$).}

\bibitem[{Pei()}]{PeierlsCoupling}
\bibinfo{note}{See, for example, K. Yonemitsu and N. Maeshima, Phys. Rev. B
  ${\mathbf{76}}$, $075105$ ($2007$); V. M. Stojanovi\'{c} and M. Vanevi\'{c},
  {\em ibid.} ${\mathbf{78}}$, $214301$ ($2008$).}

\bibitem[{\citenamefont{Park et~al.}(2007)\citenamefont{Park, Giustino, Cohen,
  and Louie}}]{ParkZgraphene:07}
\bibinfo{author}{\bibfnamefont{C.-H.} \bibnamefont{Park}},
  \bibinfo{author}{\bibfnamefont{F.}~\bibnamefont{Giustino}},
  \bibinfo{author}{\bibfnamefont{M.~L.} \bibnamefont{Cohen}}, \bibnamefont{and}
  \bibinfo{author}{\bibfnamefont{S.~G.} \bibnamefont{Louie}},
  \bibinfo{journal}{Phys. Rev. Lett.} \textbf{\bibinfo{volume}{99}},
  \bibinfo{pages}{086804} (\bibinfo{year}{2007}).

\bibitem[{Gra({\natexlab{c}})}]{GraphenePhononEffect}
\bibinfo{note}{J. L. Ma\~{n}es, Phys. Rev. B ${\mathbf{76}}$, 045430 (2007); M.
  Calandra and F. Mauri, {\em ibid.} ${\mathbf{76}}$, 205411 (2007); T.
  Stauber, N. M. R. Peres, and A. H. Castro\:Neto, {\em ibid.} ${\mathbf{78}}$,
  085418 (2008).}

\bibitem[{\citenamefont{Bostwick et~al.}(2007)\citenamefont{Bostwick, Ohta,
  Seyller, Horn, and Rotenberg}}]{Bostwick++:07}
\bibinfo{author}{\bibfnamefont{A.}~\bibnamefont{Bostwick}},
  \bibinfo{author}{\bibfnamefont{T.}~\bibnamefont{Ohta}},
  \bibinfo{author}{\bibfnamefont{T.}~\bibnamefont{Seyller}},
  \bibinfo{author}{\bibfnamefont{K.}~\bibnamefont{Horn}}, \bibnamefont{and}
  \bibinfo{author}{\bibfnamefont{E.}~\bibnamefont{Rotenberg}},
  \bibinfo{journal}{Nature Phys.} \textbf{\bibinfo{volume}{3}},
  \bibinfo{pages}{36} (\bibinfo{year}{2007}).

\bibitem[{\citenamefont{Hwang et~al.}(2007)\citenamefont{Hwang,
  Ben\:Yu-Kuang\:Hu, and Das\:Sarma}}]{Hwang++:07}
\bibinfo{author}{\bibfnamefont{E.~H.} \bibnamefont{Hwang}},
  \bibinfo{author}{\bibnamefont{Ben\:Yu-Kuang\:Hu}}, \bibnamefont{and}
  \bibinfo{author}{\bibfnamefont{S.}~\bibnamefont{Das\:Sarma}},
  \bibinfo{journal}{Phys. Rev. B} \textbf{\bibinfo{volume}{76}},
  \bibinfo{pages}{115434} (\bibinfo{year}{2007}).

\bibitem[{\citenamefont{Vukmirovi\'{c}
  et~al.}(2010)\citenamefont{Vukmirovi\'{c}, Stojanovi\'{c}, and
  Vanevi\'{c}}}]{Vukmirovic+:10}
\bibinfo{author}{\bibfnamefont{N.}~\bibnamefont{Vukmirovi\'{c}}},
  \bibinfo{author}{\bibfnamefont{V.~M.} \bibnamefont{Stojanovi\'{c}}},
  \bibnamefont{and}
  \bibinfo{author}{\bibfnamefont{M.}~\bibnamefont{Vanevi\'{c}}},
  \bibinfo{journal}{Phys. Rev. B} \textbf{\bibinfo{volume}{81}},
  \bibinfo{pages}{041408(R)} (\bibinfo{year}{2010}).

\bibitem[{\citenamefont{Saito et~al.}(1998{\natexlab{a}})\citenamefont{Saito,
  Takeya, Kimura, Dresselhaus, and Dresselhaus}}]{Saito+:98}
\bibinfo{author}{\bibfnamefont{R.}~\bibnamefont{Saito}},
  \bibinfo{author}{\bibfnamefont{T.}~\bibnamefont{Takeya}},
  \bibinfo{author}{\bibfnamefont{T.}~\bibnamefont{Kimura}},
  \bibinfo{author}{\bibfnamefont{G.}~\bibnamefont{Dresselhaus}},
  \bibnamefont{and} \bibinfo{author}{\bibfnamefont{M.~S.}
  \bibnamefont{Dresselhaus}}, \bibinfo{journal}{Phys. Rev. B}
  \textbf{\bibinfo{volume}{57}}, \bibinfo{pages}{4145}
  (\bibinfo{year}{1998}{\natexlab{a}}).

\bibitem[{Gra({\natexlab{d}})}]{GraphenePhononCalcL}
\bibinfo{note}{J.-A. Yan, W. Y. Ruan, and M. Y. Chou, Phys. Rev. B
  ${\mathbf{77}}$, $125401$ (2008); V. K. Tewary and B. Yang, {\em ibid.}
  ${\mathbf{79}}$, 075442 (2009); S. Viola\:Kusminskiy, D. K. Campbell, and A.
  H. Castro\:Neto, {\em ibid.} ${\mathbf{80}}$, 035401 (2009).}

\bibitem[{\citenamefont{Zimmermann et~al.}(2008)\citenamefont{Zimmermann,
  Pavone, and Cuniberti}}]{Zimmermann+:08}
\bibinfo{author}{\bibfnamefont{J.}~\bibnamefont{Zimmermann}},
  \bibinfo{author}{\bibfnamefont{P.}~\bibnamefont{Pavone}}, \bibnamefont{and}
  \bibinfo{author}{\bibfnamefont{G.}~\bibnamefont{Cuniberti}},
  \bibinfo{journal}{Phys. Rev. B} \textbf{\bibinfo{volume}{78}},
  \bibinfo{pages}{045410} (\bibinfo{year}{2008}).

\bibitem[{\citenamefont{Perebeinos and Tersoff}(2009)}]{Perebeinos+Tersoff:09}
\bibinfo{author}{\bibfnamefont{V.}~\bibnamefont{Perebeinos}} \bibnamefont{and}
  \bibinfo{author}{\bibfnamefont{J.}~\bibnamefont{Tersoff}},
  \bibinfo{journal}{Phys. Rev. B} \textbf{\bibinfo{volume}{79}},
  \bibinfo{pages}{241409(R)} (\bibinfo{year}{2009}).

\bibitem[{Bre()}]{Brenner+Tersoff}
\bibinfo{note}{J. Tersoff, Phys. Rev. Lett. ${\mathbf{61}}$, 2879 ($1988$);
  Phys. Rev. B ${\mathbf{37}}$, $6991$ ($1988$); D. W. Brenner, {\em ibid.}
  ${\mathbf{42}}$, $9458$ ($1990$).}

\bibitem[{\citenamefont{Jishi et~al.}(1993)\citenamefont{Jishi, Venkataraman,
  Dresselhaus, and Dresselhaus}}]{Jishi4nnfc:93}
\bibinfo{author}{\bibfnamefont{R.~A.} \bibnamefont{Jishi}},
  \bibinfo{author}{\bibfnamefont{L.}~\bibnamefont{Venkataraman}},
  \bibinfo{author}{\bibfnamefont{M.~S.} \bibnamefont{Dresselhaus}},
  \bibnamefont{and}
  \bibinfo{author}{\bibfnamefont{G.}~\bibnamefont{Dresselhaus}},
  \bibinfo{journal}{Chem. Phys. Lett.} \textbf{\bibinfo{volume}{209}},
  \bibinfo{pages}{77} (\bibinfo{year}{1993}).

\bibitem[{\citenamefont{Saito et~al.}(1998{\natexlab{b}})\citenamefont{Saito,
  Dresselhaus, and Dresselhaus}}]{CNTbook}
\bibinfo{author}{\bibfnamefont{R.}~\bibnamefont{Saito}},
  \bibinfo{author}{\bibfnamefont{G.}~\bibnamefont{Dresselhaus}},
  \bibnamefont{and} \bibinfo{author}{\bibfnamefont{M.~S.}
  \bibnamefont{Dresselhaus}}, \emph{\bibinfo{title}{Physical {P}roperties of
  {C}arbon {N}anotubes}} (\bibinfo{publisher}{Imperial {C}ollege {P}ress},
  \bibinfo{address}{London}, \bibinfo{year}{1998}{\natexlab{b}}).

\bibitem[{SSH({\natexlab{b}})}]{SSHnanotubes}
\bibinfo{note}{V. Perebeinos, J. Tersoff, and Ph. Avouris, Phys. Rev. Lett.
  ${\mathbf{94}}$, 086802 (2005); {\em ibid.} ${\mathbf{94}}$, 027402 (2005);
  L. E. F. Foa Torres and S. Roche, {\em ibid.} ${\mathbf{97}}$, 076804
  (2006).}

\bibitem[{\citenamefont{Borysenko et~al.}(2010)\citenamefont{Borysenko, Mullen,
  Barry, Paul, Semenov, Zavada, Buongiorno\:Nardelli, and
  Kim}}]{Borysenko++:10}
\bibinfo{author}{\bibfnamefont{K.~M.} \bibnamefont{Borysenko}},
  \bibinfo{author}{\bibfnamefont{J.~T.} \bibnamefont{Mullen}},
  \bibinfo{author}{\bibfnamefont{E.~A.} \bibnamefont{Barry}},
  \bibinfo{author}{\bibfnamefont{S.}~\bibnamefont{Paul}},
  \bibinfo{author}{\bibfnamefont{Y.~G.} \bibnamefont{Semenov}},
  \bibinfo{author}{\bibfnamefont{J.~M.} \bibnamefont{Zavada}},
  \bibinfo{author}{\bibfnamefont{M.}~\bibnamefont{Buongiorno\:Nardelli}},
  \bibnamefont{and} \bibinfo{author}{\bibfnamefont{K.~W.} \bibnamefont{Kim}},
  \bibinfo{journal}{Phys. Rev. B} \textbf{\bibinfo{volume}{81}},
  \bibinfo{pages}{121412} (\bibinfo{year}{2010}).

\bibitem[{\citenamefont{Holstein}(1959)}]{Holstein:59}
\bibinfo{author}{\bibfnamefont{T.}~\bibnamefont{Holstein}},
  \bibinfo{journal}{Ann. Phys.} \textbf{\bibinfo{volume}{8}},
  \bibinfo{pages}{343} (\bibinfo{year}{1959}).

\bibitem[{Dev()}]{DevreeseAlexandrovReview}
\bibinfo{note}{For a recent review, see J. T. Devreese and A. S. Alexandrov,
  Rep. Prog. Phys. ${\mathbf{72}}$, 066501 (2009).}

\bibitem[{zol()}]{zolicouple}
\bibinfo{note}{M. Zoli, Phys. Rev. B ${\mathbf{66}}$, $012303$ ($2002$);
  ${\mathbf{67}}$, $195102$ ($2003$); ${\mathbf{70}}$, $184301$ ($2004$);
  ${\mathbf{71}}$, $205111$ ($2005$).}

\bibitem[{\citenamefont{Gunnarsson}(1997)}]{Gunnarsson:97}
\bibinfo{author}{\bibfnamefont{O.}~\bibnamefont{Gunnarsson}},
  \bibinfo{journal}{Rev. Mod. Phys.} \textbf{\bibinfo{volume}{69}},
  \bibinfo{pages}{575} (\bibinfo{year}{1997}).

\bibitem[{Gun()}]{Gunnarsson+Roesch}
\bibinfo{note}{O. Gunnarsson and O. R{\"{o}}sch, Phys. Rev. B ${\mathbf{73}}$,
  174521 ($2006$); J. Phys.: Condens. Matter ${\mathbf{20}}$, 043201 (2008).}

\bibitem[{Vuk()}]{Vukmirovic+:07}
\bibinfo{note}{See, for example, N. Vukmirovi\'{c}, Z. Ikoni\'{c}, D. Indjin,
  and P. Harrison, Phys. Rev. B ${\mathbf{76}}$, $245313$ ($2007$).}

\bibitem[{Dam()}]{DamascelliARPES:04}
\bibinfo{note}{For a review, see A. Damascelli, Phys. Scr. $\mathbf{109}$, 61
  ($2004$).}

\bibitem[{Kir()}]{Kirkegaard+:05}
\bibinfo{note}{See, e.g., C. Kirkegaard, T. K. Kim, and Ph. Hofmann, New J.
  Phys. ${\mathbf{7}}$, $99$ ($2005$); T. Y. Chien, E. D. L. Rienks, M.
  Fuglsang Jensen, P. Hofmann, and E. W. Plummer, Phys. Rev. B ${\mathbf{80}}$,
  241416(R) (2009).}

\bibitem[{\citenamefont{Mahan}(1990)}]{MahanBook}
\bibinfo{author}{\bibfnamefont{G.~D.} \bibnamefont{Mahan}},
  \emph{\bibinfo{title}{Many-{P}article {P}hysics}} (\bibinfo{publisher}{Plenum
  {P}ress}, \bibinfo{address}{New York}, \bibinfo{year}{1990}).

\bibitem[{Han()}]{Hannewald+Hatch+Koller}
\bibinfo{note}{K. Hannewald, V. M. Stojanovi\'{c}, and P. A. Bobbert, J. Phys.:
  Condens. Matter ${\mathbf {16}}$, $2023$ ($2004$); K. Hannewald, V. M.
  Stojanovi\'{c}, J. M. T. Schellekens, P. A. Bobbert, G. Kresse, and J.
  Hafner, Phys. Rev. B ${\mathbf{69}}$, $075211$ ($2004$); R. C. Hatch, D. L.
  Huber, and H. H\"{o}chst, {\em ibid.} ${\mathbf{80}}$, $081411$ ($2009$);
  Phys. Rev. Lett. ${\mathbf{104}}$, 047601 (2010).}

\bibitem[{\citenamefont{Lazzeri et~al.}(2008)\citenamefont{Lazzeri,
  Attaccalite, Wirtz, and Mauri}}]{Lazzeri+:08}
\bibinfo{author}{\bibfnamefont{M.}~\bibnamefont{Lazzeri}},
  \bibinfo{author}{\bibfnamefont{C.}~\bibnamefont{Attaccalite}},
  \bibinfo{author}{\bibfnamefont{L.}~\bibnamefont{Wirtz}}, \bibnamefont{and}
  \bibinfo{author}{\bibfnamefont{F.}~\bibnamefont{Mauri}},
  \bibinfo{journal}{Phys. Rev. B} \textbf{\bibinfo{volume}{78}},
  \bibinfo{pages}{081406} (\bibinfo{year}{2008}).

\bibitem[{\citenamefont{Capone et~al.}(1997)\citenamefont{Capone, Stephan, and
  Grilli}}]{Capone+:97}
\bibinfo{author}{\bibfnamefont{M.}~\bibnamefont{Capone}},
  \bibinfo{author}{\bibfnamefont{W.}~\bibnamefont{Stephan}}, \bibnamefont{and}
  \bibinfo{author}{\bibfnamefont{M.}~\bibnamefont{Grilli}},
  \bibinfo{journal}{Phys. Rev. B} \textbf{\bibinfo{volume}{56}},
  \bibinfo{pages}{4484} (\bibinfo{year}{1997}).

\bibitem[{\citenamefont{Roche et~al.}(2005)\citenamefont{Roche, Jiang, Triozon,
  and Saito}}]{Roche+:05}
\bibinfo{author}{\bibfnamefont{S.}~\bibnamefont{Roche}},
  \bibinfo{author}{\bibfnamefont{J.}~\bibnamefont{Jiang}},
  \bibinfo{author}{\bibfnamefont{F.}~\bibnamefont{Triozon}}, \bibnamefont{and}
  \bibinfo{author}{\bibfnamefont{R.}~\bibnamefont{Saito}},
  \bibinfo{journal}{Phys. Rev. Lett.} \textbf{\bibinfo{volume}{95}},
  \bibinfo{pages}{076803} (\bibinfo{year}{2005}).

\bibitem[{And()}]{Ando++nanotube}
\bibinfo{note}{H. Suzuura and T. Ando, Phys. Rev. B ${\mathbf{65}}$, $235412$
  ($2002$); T. Ando, J. Phys. Soc. Jpn. ${\mathbf{74}}$, $777$ ($2005$).}

\bibitem[{\citenamefont{Fratini and Guinea}(2008)}]{Fratini:08}
\bibinfo{author}{\bibfnamefont{S.}~\bibnamefont{Fratini}} \bibnamefont{and}
  \bibinfo{author}{\bibfnamefont{F.}~\bibnamefont{Guinea}},
  \bibinfo{journal}{Phys. Rev. B} \textbf{\bibinfo{volume}{77}},
  \bibinfo{pages}{195415} (\bibinfo{year}{2008}).

\bibitem[{DeF()}]{DeFilippis+:10}
\bibinfo{note}{G. De Filippis, V. Cataudella, S. Fratini, and S. Ciuchi,
  e-print arXiv:1005.2476 (2010).}

\bibitem[{Ber({\natexlab{b}})}]{BerciuMomDependent}
\bibinfo{note}{C. Slezak, A. Macridin, G. A. Sawatzky, M. Jarrell, and T. A.
  Maier, Phys. Rev. B ${\mathbf{73}}$, 205122 (2006); B. Lau, M. Berciu, and G.
  A. Sawatzky, {\em ibid.} ${\mathbf{76}}$, $174305$ ($2007$); G. L. Goodvin
  and M. Berciu, {\em ibid.} ${\mathbf{78}}$, $235120$ ($2008$).}

\bibitem[{Gra({\natexlab{e}})}]{GrapheneMetalCont}
\bibinfo{note}{S. Barraza-Lopez, M. Vanevi\'{c}, M. Kindermann, and M. Y. Chou,
  Phys. Rev. Lett. ${\mathbf{104}}$, $076807$ ($2010$).}

\end{thebibliography}
\bibliographystyle{apsrev}
\end{document}